\documentclass[aps,pra,showpacs,twocolumn,eqsecnum,superscriptaddress]{revtex4}

\usepackage{amsmath}
\usepackage{amssymb}
\usepackage{latexsym}
\usepackage{graphicx}

\newcommand{\non}{\nonumber}
\newcommand{\ssize}{\scriptsize}

\newcommand{\ie}{i.e.\ }
\newcommand{\etal}{{\em et al.}}

\newcommand{\tr}{\ensuremath{\mathrm{tr}}}
\newcommand{\Sch}{\ensuremath{\mathrm{Sch}}}
\renewcommand{\>}{\ensuremath{\rangle}}
\newcommand{\<}{\ensuremath{\langle}}
\newcommand{\ket}{\ensuremath{\rangle}}
\newcommand{\bra}{\ensuremath{\langle}}
\newcommand{\tp}{\ensuremath{\otimes}}
\renewcommand{\l}{\left}
\renewcommand{\r}{\right}
\renewcommand{\[}{\left[}
\renewcommand{\]}{\right]}
\renewcommand{\(}{\left(}
\renewcommand{\)}{\right)}
\newcommand{\eq}{\equiv}
\newcommand{\A}{\ensuremath{{\cal A}}}
\newcommand{\B}{\ensuremath{{\cal B}}}
\newcommand{\C}{\ensuremath{{\cal C}}}
\newcommand{\D}{\ensuremath{{\cal D}}}
\newcommand{\E}{\ensuremath{{\cal E}}}
\newcommand{\F}{\ensuremath{{\cal F}}}
\newcommand{\I}{\ensuremath{{\cal I}}}
\newcommand{\LU}{\ensuremath{\mathbb{LU}}}
\newcommand{\R}{\ensuremath{{\cal R}}}
\newcommand{\U}{\ensuremath{\mathbb{U}}}

\newcommand{\Kdel}{\ensuremath{K_{\Delta E}}}
\newcommand{\Khar}{\ensuremath{K_\mathrm{Har}}}
\newcommand{\Ke}{\ensuremath{K_E}}
\newcommand{\Ksch}{\ensuremath{K_\mathrm{Sch}}}
\newcommand{\Kd}{\ensuremath{K_D}}
\newcommand{\Khs}{\ensuremath{K_\mathrm{HS}}}
\newcommand{\Kcom}{\ensuremath{K_\mathrm{com}}}
\newcommand{\Kmax}{\ensuremath{K_{\max}}}

\newcommand{\cnot}{\textsc{cnot}}
\newcommand{\swap}{\textsc{swap}}

\newtheorem{theorem}{Theorem}
\newtheorem{lemma}[theorem]{Lemma}
\newtheorem{proposition}[theorem]{Proposition}
\newcommand{\proof}{\noindent{\bf Proof:} }
\newcommand{\qed}{\hfill $\blacksquare$}
\newtheorem{axiom}{Axiom}
\newtheorem{property}{Property}

\begin{document}

\title{Quantum dynamics as a physical resource}

\author{Michael~A.~Nielsen} \email{nielsen@physics.uq.edu.au}
\author{Christopher~M.~Dawson}
\author{Jennifer~L.~Dodd} \email{jdodd@physics.uq.edu.au}
\author{Alexei~Gilchrist}
\author{Duncan~Mortimer}
\author{Tobias~J.~Osborne}
\author{Michael~J.~Bremner}
\affiliation{Centre for Quantum Computer Technology and Department of
Physics, The University of Queensland, Brisbane 4072 Australia}
\author{Aram~W.~Harrow}
\affiliation{MIT Physics, 77 Massachusetts Ave., Cambridge MA 02139
USA}
\author{Andrew~Hines}
\affiliation{Centre for Quantum Computer Technology and Department of
Physics, The University of Queensland, Brisbane 4072 Australia}
\date{\today}

\begin{abstract}
How useful is a quantum dynamical operation for quantum information
processing?  Motivated by this question we investigate several {\em
strength measures} quantifying the resources intrinsic to a quantum
operation.  We develop a general theory of such strength measures,
based on axiomatic considerations independent of state-based
resources.  The power of this theory is demonstrated with applications
to quantum communication complexity, quantum computational complexity,
and entanglement generation by unitary operations.
\end{abstract}


\maketitle


%
%
\section{Introduction}
%
%
The quantification and comparison of different types of physical
resources lies at the heart of much of modern science.  A good example
is the physical resource {\em energy}, whose quantification enabled
the development of thermodynamics.  More recently, motivated by
applications to quantum information processing, there have been
attempts to develop a quantitative theory of quantum
entanglement~\cite{Bennett96c}.  This theory, still in its nascent
stages, has been applied to gain insight into questions about the
capacity of a noisy channel for information~\cite{Schumacher02},
quantum teleportation with a noisy entangled
resource~\cite{Horodecki99d}, and distributed quantum
computation~\cite{Nielsen98}.

%
%
Structurally, quantum mechanics has two parts, one part concerned with
{\em quantum states}, the other with {\em quantum dynamics}.  A
general quantum dynamical process is described by a {\em quantum
  operation} (reviewed in~\cite{Nielsen00}); such processes include
unitary evolution, quantum measurement, dissipation, and decoherence.
We believe quantum operations are a useful physical resource on an
equal and logically independent footing to quantum states.

%
%
The first step in studying a physical resource is to quantify it.
Therefore, the purpose of our paper is to develop a theory quantifying
the {\em strength} of quantum dynamical operations.  Our motivations
are axiomatic and operational questions concerning quantum dynamics.
Our goal is to find strength measures capturing some of the structure
in the complicated space of quantum operations, to gain insight into
quantum dynamics and complex quantum
systems~\cite{Nielsen02,Osborne02}.  Although some of the measures we
propose for operations are based on state entanglement measures, we
expect the study of dynamics to provide different, complementary
insights to those gained from the study of states.

%
%
What questions will good strength measures allow us to analyze?  We
foresee applications to the analysis of quantum computational
complexity, distributed quantum computation, quantum communication,
and quantum cryptography.  As a simple example, consider the question
of how many controlled-\textsc{not} (\cnot) gates are required to
implement a \swap\ gate on two qubits, when assisted by arbitrary
local unitaries.  Suppose we have a measure $K(U)$, quantifying the
strength of a unitary $U$.  Suppose further that $K(U)$ satisfies (a)
$K(UV)\le K(U)+K(V)$; and (b) $K(U)=0$ for local unitaries $U$.  It is
easy to see that the number of $\cnot$ gates needed to do the \swap\ 
gate is at least $K(\swap)/K(\cnot)$.

%
%
More generally, the central problem of quantum computational
complexity is to determine the minimum number of one- and two-qubit
gates necessary to implement a desired $n$-qubit unitary operation
$U$.  For example, $U$ might encode the solution to a problem such as
the traveling salesman problem.  Suppose we have a strength measure
satisfying properties (a) and (b), above, as well as (c) $K(U\tp
I)=K(U)$.  The number of gates needed to compute $U$ is again bounded
below by $K(U)/K(\cnot)$.  Such a bound might help in determining the
relationships between various quantum and classical complexity
classes.  We will return to this application several times.

Another motivation to study quantum dynamics as a resource is recent
work on {\em universality} in quantum computation.  The class of
interactions capable of performing universal quantum computation has
been shown to be the class of bipartite entangling dynamics; any
Hamiltonian which can create entanglement between any pair of qudits
is universal, when assisted by arbitrary one-qudit unitaries
(see~\cite{Dodd02,Wocjan02,Dur01,Bennett01,Nielsen01b,Vidal01b} and
references therein, see also~\cite{Jones99,Leung00} for related work).
It has also been shown that any entangling two-qudit unitary, together
with arbitrary one-qudit unitaries, is universal (\cite{Brylinski01},
see~\cite{Bremner02} for a simple, constructive proof in the qubit
case).

These results show that there is a qualitative difference between
entangling and non-entangling dynamics.  Furthermore, they show all
two-qudit entangling dynamics are qualitatively equivalent, as any one
can simulate any other, provided local unitaries are available.  By
analogy with the study of state entanglement, this suggests
quantifying entangling dynamics. We now review prior work on this
idea, organizing our discussion around three motivating themes: the
communication cost to implement an operation; the entangling ability
of an operation; and the ability of an operation to communicate bits.

%
%
The communication requirements for implementing a general bipartite
unitary $U$ were studied in Ch.\,6 of~\cite{Nielsen98}, where a
general lower bound on the number of qubits of communication needed to
implement $U$ was proved, depending only on the operator Schmidt
decomposition of $U$ (see Sec.\,\ref{se:invitation} for discussion of
this decomposition).  Eisert \etal~\cite{Eisert00b} and Collins,
Linden, and Popescu~\cite{Collins00} studied the classical
communication and entanglement required to implement some specific
few-qubit quantum gates. Chefles, Gilson, and Barnett~\cite{Chefles00}
studied the amount of communication and entanglement required to
perform an arbitrary gate in a network of qubits.

%
%
The capacity of a quantum operation to generate entanglement seems to
have first been studied by Makhlin~\cite{Makhlin00}, who found three
invariants characterizing the non-local properties of two-qubit
unitaries.  Makhlin used the invariants to obtain results about
entanglement generation, with a view towards applying them to the
complexity of implementing gates.  Zanardi, Zalka, and
Faoro~\cite{Zanardi00}, Zanardi~\cite{Zanardi01}, and Wang and
Zanardi~\cite{Wang02}, all obtained results about the {\em average
  entanglement} generated by a unitary.  Cirac \etal~\cite{Cirac01}
studied the ability of an operation to produce entanglement by mapping
the operation onto a corresponding state, and studying the properties
of that state.  Kraus and Cirac~\cite{Kraus01} studied the {\em
  maximum entanglement} which can be created by a unitary operator
acting on two initially unentangled qubits.  They found an explicit
formula for the maximum entanglement that can be generated without
ancillas, and showed that this amount can be exceeded with the use of
ancillas.  Leifer, Henderson, and Linden~\cite{Leifer02} used similar
reasoning to obtain an explicit formula for the entanglement generated
without ancillas, but allowing initial entanglement.  They also
obtained numerical results demonstrating that the addition of ancillas
can increase the maximum entanglement generated.  In a different
context, Scarani \emph{et al}~\cite{Scarani02} related the entangling
power of a unitary operation to the problem of thermalization of a
quantum system.

%
%
A related approach is to quantify the entangling abilities of
Hamiltonians rather than unitaries.  D\"ur \etal~\cite{Dur01}
considered the {\em rate} at which a Hamiltonian creates entanglement,
and found techniques to optimize this rate.  More recently, Vidal,
Hammerer, and Cirac~\cite{Vidal02} (see also~\cite{Hammerer02})
analytically characterized the minimum time required to simulate one
Hamiltonian with another, and found the minimum time required to
simulate a desired unitary with a Hamiltonian.  This allowed them to
define a partial order on unitaries, according to which one unitary
$U$ is more {\em non-local} than another unitary $V$ if and only if,
for any Hamiltonian, the minimum time required to simulate $U$ is
longer than the minimum time to simulate $V$.  They also obtained
results on the optimal choices of non-local interactions for
transmitting classical bits between two parties.  Childs
\etal~\cite{Childs02} found an explicit formula for the maximum
entanglement created by a class of two-qubit Hamiltonians, including
the Ising interaction and the anisotropic Heisenberg interaction, for
which this maximum is achieved without ancillas.

%
%
The ability of a quantum operation to communicate classical
information was studied by Beckman \etal~\cite{Beckman01}, who
obtained simple necessary and sufficient conditions for information
transmission to be possible.  Bennett \etal~\cite{Bennett02} and Berry
and Sanders~\cite{Berry02b} studied the {\em capacity} of a bipartite
operation to communicate information, and related this capacity to
the ability of the operation to generate entanglement.

%
%
Our paper draws on all these perspectives, but differs in an important
way.  Rather than focusing on the ability of a quantum dynamical
operation to generate some static resource, such as entanglement or
shared classical bits, we believe it is possible to quantify quantum
dynamical operations as a physical resource in their own right.  That
is, we do not need to make reference to the ability of the operation
to generate some other resource.

%
%
How can one develop a theory of dynamic strength without relying on
familiar state-based resources?  The approach we take is to identify
plausible axioms and properties a good measure of strength should
satisfy, and develop measures satisfying those properties.

%
%
The paper is structured as follows.  Sec.\,\ref{se:invitation} opens
by introducing two concrete examples of strength measures for unitary
operations, the {\em Hartley} and {\em Schmidt} strengths.
Sec.\,\ref{se:framework} considers operational questions motivating
strength measures, and uses these questions to motivate some abstract
axioms for such measures.  Sec.\,\ref{se:interlude} briefly summarizes
a useful {\em canonical decomposition} for two-qubit unitary
operators.  Secs.\,\ref{se:generation} and~\ref{se:metrics} explore a
variety of specific definitions for dynamic strength measures.  Our
general philosophy is to explore a wide variety of measures, and then
to concentrate on those which appear most likely to yield useful
practical answers to interesting operational questions.
Sec.\,\ref{se:future} concludes with a summary and a table of results.



\section{Invitation: the Hartley and Schmidt strengths}
\label{se:invitation}

In this section we introduce two strength measures, the {\em Hartley
  strength} and the {\em Schmidt strength}.  These measures are
introduced both because of their intrinsic interest, and also because
the Hartley strength will be used as a simple, concrete example
illustrating the more abstract, axiomatic approach to dynamic
strength.  The Hartley and Schmidt strengths are based on a
generalization of the Schmidt decomposition to operators, which we
call the operator-Schmidt decomposition~\cite{Nielsen98}.  To explain
the operator-Schmidt decomposition we introduce the Hilbert-Schmidt
inner product on the space of $d\times d$ operators, defined by
$(Q,P)\eq\tr(Q^\dag P)$, for any operators $Q$ and $P$.  Using this
inner product, we define an orthonormal operator basis to be a set
$\{Q_j\}$ which satisfies the condition: $(Q_j,Q_k)=\tr(Q_j^\dag
Q_k)=\delta_{jk}$ where $\delta_{jk}=1$ if $j=k$ and 0 otherwise.  For
example, a complete orthonormal basis for the space of one-qubit
operators is the set of normalized Pauli matrices, $\{\tilde I,\tilde
X,\tilde Y,\tilde Z\}\eq\{I/\sqrt2,X/\sqrt2,Y/\sqrt2,Z/\sqrt2\}$.

%
%
An operator $Q$ acting on systems \A\ and \B\ may be written in the
operator-Schmidt decomposition~\cite{Nielsen98}:
\begin{equation} \label{eq:operator_schmidt}
Q=\sum\nolimits_l s_l A_l\tp B_l,
\end{equation}
where $s_l\ge0$ and $A_l$ and $B_l$ are orthonormal operator bases for
\A\ and \B, respectively.  To prove the operator-Schmidt
decomposition, expand $Q=\sum_{jk}M_{jk}C_j\tp D_k$, where $C_j$ and
$D_k$ are fixed orthonormal operator bases for \A\ and \B,
respectively, and $M_{jk}$ are coefficients.  The singular value
decomposition states that the matrix $M$, whose $(j,k)$th entry is
$M_{jk}$, may be written $M=UsV$, where $U$ and $V$ are unitary, and
$s$ is diagonal with non-negative entries.  We thus obtain
\begin{equation}
Q=\sum_{jkl}U_{jl}s_l V_{lk}C_j\tp D_k,
\end{equation}
where $s_l$ is the $l$th diagonal entry of $s$.  Defining
$A_l\eq\sum_j U_{jl}C_j$ and $B_l\eq\sum_k V_{lk}D_k$, which are
easily shown to be orthonormal operator bases for \A\ and \B, we
obtain the operator-Schmidt decomposition
Eq.\,(\ref{eq:operator_schmidt}).

%
%
Nielsen~\cite{Nielsen98} defines the Schmidt number of an operator,
$\Sch(Q)$, to be the number of non-zero coefficients in the
operator-Schmidt decomposition for $Q$~\footnote{Terhal and
  Horodecki~\cite{Terhal00c} defined an alternative notion of
  Schmidt number for bipartite density matrices.}.

A simple example is the \cnot\ gate which has operator-Schmidt
decomposition
\begin{equation}
\cnot=\sqrt2\,|0\>\<0|\tp\tilde{I}+\sqrt2\,|1\>\<1|\tp
\tilde{X}
\end{equation}
and hence has Schmidt coefficients $\{\sqrt2,\sqrt2\}$, and
$\Sch(\cnot)=2$.  The \swap\ gate for qubits has operator-Schmidt
decomposition
\begin{equation}
\swap=\tilde{I}\tp\tilde{I}+\tilde{X}\tp\tilde{X}+\tilde{Y}\tp
\tilde{Y}+\tilde{Z}\tp\tilde{Z}
\end{equation}
and hence $\Sch(\swap)=4$.  A less familiar example is the gate
\begin{equation} \begin{split} \label{eq:U_p}
U_p=\(\sqrt{1-p}\,I\tp I+i\sqrt{p}\,X\tp X\)\times\\
\(\sqrt{1-p}\,I\tp I+i\sqrt{p}\,Z\tp Z\)
\end{split} \end{equation}
which has operator-Schmidt decomposition
\begin{equation} \begin{split}
U_p=&\ 2\(1-p\)\tilde{I}\tp\tilde{I}+2p\tilde{Y}\tp\tilde{Y}+\\
&\ 2\sqrt{p(1-p)}\[\(e^{i\pi/4}\tilde{X}\)\tp\tilde{X}+\(e^{i\pi/4}
\tilde{Z}\)\tp\tilde{Z}\]
\end{split} \end{equation}
and thus has Schmidt number 1 when $p=0$ or 1, and 4 otherwise.

%
%
A more complicated example is provided by the quantum Fourier
transform, whose unitary action on $l$ qubits is defined by the action
on computational basis states~\cite{Shor94}
\begin{equation}
|s\>\to\frac{1}{\sqrt{2^l}}\sum_{t=0}^{2^l-1}e^{2\pi ist/2^l}|t\>,
\end{equation}
where we number the basis states from $|0\>$ through $|2^l-1\>$.  A
useful alternate formula for the quantum Fourier transform may be
obtained by working in a binary representation, $s=s_1\cdots s_l$,
whence~\footnote{According to Chapter~12 of~\cite{Press92}, this
  decomposition was obtained by Danielson and Lanczos in 1942.  It was
  rediscovered in the quantum context in~\cite{Griffiths96}.}
\begin{equation} \label{eq:DLQFT}
|s_1,\ldots,s_l\>\to|f_{s_l}\>\tp |f_{s_{l-1}s_l}\>\tp\cdots|f_{s_1
\cdots s_l}\>,
\end{equation}
where the one-qubit state $|f_t\>$ is defined for an arbitrary bit
string $t=t_1\cdots t_k$ by $|f_t\>\eq[|0\>+\exp(2\pi
i\,0.t)|1\>]/\sqrt2$, and $0.t$ is the binary fraction
$t_1/2+t_2/4+\cdots+t_k/2^k$.

%
%
Suppose now that system \A\ consists of $m$ qubits and system \B\
consists of $n$ qubits, and $U$ is the quantum Fourier transform on
$m+n$ qubits.  From Eq.\,(\ref{eq:DLQFT}),
\begin{equation}
U|x_1,\ldots,x_m,y_1,\ldots,y_n\>=|f_{y_n}\>\tp\cdots\tp|f_{x_1\cdots
y_n}\>.
\end{equation}
Suppose $m\le n$.  To determine the Schmidt decomposition of the
quantum Fourier transform it is convenient to introduce the notation
$y'=y_1\cdots y_{n-m}$ and $y''=y_{n-m+1}\cdots y_n$, so the string
$y$ can be formed by concatenating the strings $y'$ and $y''$.  It
follows from the previous equation that
\begin{equation}
U=\sum_{xy''}A_{xy''}\tp B_{xy''},
\end{equation}
where $x$ ranges over $m$-bit strings $x_1\cdots x_m$, and we define
\begin{equation} \begin{split}
A_{xy''}&\eq|f_{y_n}\>|f_{y_{n-1}y_n}\>\cdots|f_{y_{n-m+1}\cdots y_n}
\>\< x|\\
B_{xy''}&\eq\sum_{y'}C_{xy'y''}\\
C_{xy'y''}&\eq|f_{y_{n-m}\cdots y_n}\>\cdots|f_{x_1\cdots y_n}\>\<y|.
\end{split} \end{equation}
A calculation shows that the $A_{xy''}$ are orthonormal operators, and
the $B_{xy''}$ are an orthogonal set, with
$(B_{xy''},B_{xy''})=2^{n-m}$.  Thus the Schmidt decomposition for the
quantum Fourier transform is
\begin{equation}
U=\sum_{xy''}\sqrt{2^{n-m}}A_{xy''}\tp\frac{B_{xy''}}{\sqrt{2^{n-m}}}.
\end{equation}
Thus, when $m\le n$ the quantum Fourier transform has Schmidt number
$2^{2m}$, and all nonzero Schmidt coefficients are equal to
$\sqrt{2^{n-m}}$.  Note that the Schmidt decomposition of the quantum
Fourier transform was already obtained in~\cite{Nielsen98} when $m=n$;
we have not yet succeeded in determining the Schmidt decomposition of
the quantum Fourier transform when $m>n$, but conjecture that it has
Schmidt number $2^{2n}$\footnote{Tyson~\cite{Tyson02} has recently
  calculated the Schmidt decomposition of the quantum Fourier
  transform in the general case, and has confirmed this conjecture.}

%
%
The {\em Hartley strength}~\footnote{The term Hartley strength comes
from the Hartley entropy~\cite{Hartley28} of a probability
distribution, defined to be the logarithm of the number of non-zero
elements in the probability distribution.} of an operator $\Khar(Q)$
is defined by
\begin{equation}
\Khar(Q)\eq\log\[\Sch(Q)\].
\end{equation}
(The logarithm is taken to base 2 throughout this paper.)  Returning
to our examples, the \cnot\ gate has Hartley strength $\log2=1$, the
\swap\ gate has strength 2, and $U_p$ has strength 0 for $p=0$ or 1,
and strength $\log4=2$ otherwise.  The quantum Fourier transform has
Hartley strength $2m$, provided $m\le n$.

The {\em Schmidt strength} is motivated by a simple observation about
unitary operators $U$ acting on systems \A\ and \B\ of respective
dimensions $d_\A$ and $d_\B$.  For such an operator, the relation
$\tr(U^\dag U)=d_\A d_\B$ implies that the Schmidt coefficients $s_l$
satisfy $\sum_l s_l^2=d_\A d_\B$.  Therefore, the numbers $s_l^2/(d_\A
d_\B)$ form a probability distribution.  A natural measure of the
non-local content of $U$ is thus the {\em Schmidt strength}, defined
to be the Shannon entropy $H(\cdot)$ of the distribution $s_l^2/(d_\A
d_\B)$~\footnote{Note that Zanardi~\cite{Zanardi01}, and Wang and
  Zanardi~\cite{Wang02}, have investigated similar, though
  inequivalent, measures for unitaries.  This work is discussed in
  Sec.~\ref{subsubsec:ent-gen-comm}.},
\begin{equation} \label{eq:unitary-Ksch}
\Ksch(U)\eq H\textstyle{\(\l\{\frac{s_l^2}{d_\A d_\B}\r\}\)}.
\end{equation}
More generally, for an arbitrary bipartite operator $Q$ we define the
Schmidt strength by
\begin{equation} \label{eq:op-Ksch}
\Ksch(Q)\eq H\textstyle{\(\l\{\frac{s_l^2}{\tr(Q^\dag Q)}\r\}\)},
\end{equation}
where $\{s_l^2/\tr(Q^\dag Q)\}$ are the squared Schmidt coefficients
of $Q$, normalized to form a probability distribution.  Note that
$\Ksch(\cnot)=1$, $\Ksch(\swap)=2$,
$\Ksch(U_p)=H\[(1-p)^2,p^2,p(1-p),p(1-p)\]$, and $\Ksch=2m$ for the
quantum Fourier transform, when $m\le n$.


\section{Conceptual framework} \label{se:framework}

In this section, we explore two approaches to the definition of
strength measures. In the {\em operational approach}, discussed in
Sec.\,\ref{subse:operational}, we define several measures of strength
based on the ability of an operation to perform various tasks.  These
measures thus quantify a dynamical resource required by each task.
The second approach, the {\em axiomatic approach}, is explored in
Sec.\,\ref{subse:axiomatic}, where we identify a list of three axioms
and nine useful properties for a strength measure.  These two
approaches may appear to be independent, but there is actually
substantial interplay.  In particular, many of the properties in
Sec.\,\ref{subse:axiomatic} are motivated by consideration of the
operational measures of strength of Sec.\,\ref{subse:operational}.


\subsection{Operational approach} \label{subse:operational}

Quantum dynamics are clearly an essential component in quantum
information processing tasks.  However, it is difficult to identify
which properties of quantum dynamics are the most essential, because
different properties are required for different tasks.  This variety
is reflected in this section by the fact that different operational
questions give rise to different notions of strength.

%
%
The reader should note that the main point of this section is not to
prove results about the measures we define.  Rather, it is to provide
definitions of some strength measures, and a discussion of their
operational motivation.  After we have enumerated the properties we
would like these measures to satisfy, we will take up the problem of
determining the properties of these measures, and the relationships
between them.


\subsubsection{Entanglement generation and communication capacity}
\label{subsubsec:ent-gen-comm}

In this section, we consider two related questions on the ability of a
quantum operation to create entanglement and to communicate
information.  We also review some of the recent work on these
subjects.\\

\noindent{\bf How much entanglement can be generated by a quantum
operation?}

How much entanglement a single application of a unitary $U$ can
generate depends crucially on the initial states $U$ may act on.  We
must also specify whether we are interested in the maximum, minimum,
or average entanglement generated.  We focus primarily on
maximizations.

We define two measures for the {\em entangling strength} of a unitary
$U$.  (See Sec.\,\ref{subsubse:extension} for some generalizations to
quantum operations.)

The first strength measure quantifies the maximum entanglement which a
unitary $U$ can create between two systems \A\ and \B\ with the use of
arbitrary ancillas, but without prior entanglement:
\begin{equation} \label{eq:kanc}
\Ke(U)\eq\max_{|\alpha\>,|\beta\>}E(U|\alpha\>|\beta\>)
\end{equation}
where $|\alpha\>$ ranges over all (possibly entangled) states of
system \A\ plus an ancilla $\R_\A$, and $|\beta\>$ ranges over states
of system \B\ plus an ancilla $\R_\B$, and $E$ is the usual measure of
bipartite pure state entanglement, the von Neumann entropy of the
reduced density matrix~\footnote{Maximizing over mixed states as well
as pure states does not change the value of \Ke\ because of the
presence of arbitrary ancillas.  In particular, suppose \A\ and \B\
were in states $\rho_\A$ and $\rho_\B$ respectively.  By introducing
copies of their systems, $\R_\A$ and $\R_\B$, it is possible to find
pure states $|\alpha\>$ and $|\beta\>$ of $\A\R_\A$ and $\B\R_\B$ such
that $\tr_{\R_\A}(|\alpha\>\<\alpha|)=\rho_\A$ and
$\tr_{\R_\B}(|\beta\>\<\beta|)=\rho_\B$.  Since entanglement decreases
when systems are discarded, we must have $E(U|\alpha\>|\beta\>)\ge
E(U\rho_\A\tp\rho_\B U^\dag)$.}.  Note that the ancillas may be chosen
with dimensions equal to the dimensions of \A\ and \B\, respectively,
since the Schmidt number of $|\alpha\>$ with respect to the $\A:\R_\A$
division is at most $d_\A$, and similarly for $|\beta\>$. It follows
that \Ke\ is truly a maximum, and not a supremum.

Kraus and Cirac~\cite{Kraus01} calculated $\Ke(U)$ for some special
two-qubit unitaries, while Leifer, Henderson and
Linden~\cite{Leifer02} obtained numerical evidence that removing the
ancillas decreases the maximum entanglement for certain unitaries.

The second measure allows the possibility of prior entanglement as
well as ancillas.  $\Kdel(U)$ is the magnitude of the maximal {\em
change} in entanglement caused by $U$:
\begin{equation} \label{eq:kdel}
\Kdel(U)\eq\sup_{|\psi\>}\l|E(U|\psi\>)-E(|\psi\>)\r|,
\end{equation}
where $|\psi\>$ ranges over all states of $\A\R_\A$ and
$\B\R_\B$~\footnote{We have defined \Kdel\ as a supremum over pure
  states.  The simple argument showing that \Ke\ may be restricted to
  pure states does not apply here, since \Kdel\ is a difference of
  entanglement measures.  In general, if \Kdel\ is extended to mixed
  states, its value may depend on the entanglement measure used.
  Bennett \etal~\cite{Bennett02} considered several cases of this
  problem, although they were interested in the maximum {\em increase}
  in entanglement, rather than the magnitude of the change in
  entanglement.  The supremum must appear in the definition of \Kdel,
  rather than a maximization as in the definition of \Ke, since we do
  not know of any bound on the size of the ancilla.}.

Clearly, $\Ke(U)\le\Kdel(U)$ for all $U$. Later, we will see that there
exist unitaries $U$ for which $\Ke(U)\ne\Kdel(U)$, demonstrating that
these two measures capture different notions of a unitary's ability to
generate entanglement.

%
%
An alternative approach to quantifying entanglement generation has
been explored by Zanardi~\cite{Zanardi01}, and Wang and
Zanardi~\cite{Wang02}.  Zanardi~\cite{Zanardi01} defines a measure of
entanglement, $L(U)$, for a unitary operator $U$ on a $d_\A \times
d_\B$ system by the linear entropy, $L(U) \equiv 1-\sum_l
s_l^4/d_{\A}^2d_{\B}^2$, where $s_l$ are the Schmidt coefficients of
$U$.  Provided $d_\A = d_\B = d$, it can be shown
that~\cite{Zanardi01,Wang02},
\begin{equation} \begin{split}
&\int d\alpha\,d\beta\,L\[U(\alpha\tp\beta)\]\\
&=\frac{d^2}{(d+1)^2}\[L(U)+L(U\swap)-L(\swap)\],
\end{split} \end{equation}
where $d\alpha$ and $d\beta$ are the uniform, normalized, Haar
measures on the first and second qudits, respectively, the function
$L$ on the left is the measure of {\em state} entanglement based on
the linear entropy of the squared Schmidt coefficients of the state,
while the function $L$ on the right is the {\em operator} entanglement
defined by Zanardi.  This equation nicely connects the Schmidt
coefficients and the average entanglement generated by $U$.

%
%
In a similar vein, Wang and Zanardi~\cite{Wang02} define a notion of
{\em concurrence} for unitary operators with Schmidt number 2.  For a
system $\A \B$ of dimension $d_\A \times d_\B$, they define
$C(U)\eq2s_1 s_2/(d_\A d_\B)$, where $s_1$ and $s_2$ are the Schmidt
coefficients of $U$.  This definition extends the notion of
concurrence for qubits introduced by Hill and Wootters~\cite{Hill97}.
Simple algebra and the fact that $\sum_l s_l^2 = d_\A d_\B$ implies
that $C^2(U)=2L(U)$, where $L(U)$ is the measure of operator
entanglement introduced by Zanardi~\cite{Zanardi01}.

\noindent{\bf How useful is a quantum operation for communication?}

An interesting question is to determine the relationship between the
entanglement generated by a channel and its capacity to transmit
classical information between two systems.  Recently, Bennett
\etal~\cite{Bennett02} and Berry and Sanders~\cite{Berry02b} have
examined the relationship between the entangling capacity of a
two-qubit unitary and its ability to transmit information.  In
particular, Bennett \etal\ considered the maximum entanglement that
can be generated from any (possibly entangled and mixed) state with
$t$ uses of the unitary gate $U$.  They argued that the maximum
entanglement generated with $t$ uses of $U$ is just $t$ times the
maximum entanglement generated with one use of $U$, and that \Kdel\ is
an upper bound on the average number of bits which can be reliably
transmitted between \A\ and \B.


\subsubsection{Quantum computational complexity}

In this section we consider a different motivation for the study of
quantum dynamics as a resource.  Rather than considering an
operation's explicitly non-local properties (such as its ability to
create entanglement), we ask what characterizes the difficulty of
performing a quantum computation.\\

%
%
%

A reasonable measure of the {\em complexity} of implementing a unitary
$U$ with a gate set \U\ is simply the minimum number of gates from \U\
in a circuit which implements $U$.  For example, suppose we only have
the ability to implement the \cnot\ gate on two qubits, with either
acting as the control, and we wish to simulate the \swap\ gate.  In
this case we have the gate set $\U=\{\cnot_{12},\cnot_{21}\}$ where
the first subscript refers to the control qubit and the second the
target.  Since $\swap=\cnot_{12}\cnot_{21}\cnot_{12}$ (and the \swap\
gate cannot be implemented with only two \cnot\ gates), the complexity
of the \swap\ gate relative to \U\ is 3.

To generalize this idea, we define \Kcom:
\begin{equation}
\Kcom(U|\U)\eq\min\l\{\sum\nolimits_j\chi(W_j)\l|U=\prod\nolimits_j
W_j,W_j\in\U\r.\r\},
\end{equation}
where the {\em cost function} $\chi(W_j)$ is any non-negative function
that quantifies the difficulty associated with implementing $W_j$.

The circuit complexity measure has the property that, for any two
unitary operators $U$ and $V$,
\begin{equation}
\Kcom(UV|\U)\le\Kcom(U|\U)+\Kcom(V|\U),
\end{equation}
since one circuit implementing $UV$ is the concatenation of the
minimal circuits implementing $V$ and $U$ separately.  We refer to
this property as the {\em chaining property}.

In general, \Kcom\ is prohibitively difficult to calculate since it is
very hard to prove that a given circuit for $U$ is minimal.  However,
it is possible to find lower bounds on \Kcom\ as follows.  Expanding
upon the example given in the introduction, suppose $U$ is a two-qudit
unitary, and one is given the ability to perform a set of two-qudit
gates $\U=\{U_1,\ldots,U_m\}$, and local unitary operations.  What is
the minimum number of two-qudit gates required to implement $U$?
Suppose $U=(A_0\tp B_0)U_{l_1}(A_1\tp B_1)\cdots U_{l_k}(A_k\tp B_k)$,
where $A_j\tp B_j$ denotes a local unitary, and $U_{l_j}\in\U$.  Let
$K$ be any measure satisfying $K(UV)\le K(U)+K(V)$ and $K(A\tp B)=0$
for any local unitary $A\tp B$. Then
\begin{equation} \begin{split}
K(U)=&\ K[(A_0\tp B_0)U_{l_1}(A_1\tp B_1)\cdots U_{l_k}(A_k\tp B_k)]\\
\le&\ K(U_{l_1})+\cdots+K(U_{l_k})\\
\le&\ k\Kmax,
\end{split} \end{equation}
where \Kmax\ is the maximum value of $K(U_{l_j})$.  We have deduced a
useful bound on the number of gates,
\begin{equation} \label{eq:lower-bound}
k\ge\frac{K(U)}{\Kmax}.
\end{equation}
This captures the intuitively appealing notion that the number of
gates required to implement $U$ is at least equal to the total
strength of $U$, divided by the maximum strength of any of the
implementing gates.  Indeed, if we take the cost of a local unitary to
be 0 and the cost of a two-qudit gate to be 1, the argument implies
that $\Kcom(U|\U)\ge K(U)/\Kmax$.  Although this argument holds only
for two-qudit unitaries, $U$, we will extend it to $n$-qudit unitaries
after the discussion of stability properties in the next section.


\subsection{Axiomatic approach}
\label{subse:axiomatic}

%
%
One approach to quantifying entanglement is to consider axioms which
an entanglement measure ``ought'' to satisfy, and to explore the
consequences of those
axioms~\cite{Bennett96c,Vedral97,Vedral98,Vidal00b}.  While this
approach has occasionally been criticized~\cite{Nielsen01c}, it has
certainly proven fruitful.

%
%
Here we explore an analogous axiomatic approach to the study of
strength measures for quantum dynamical operations.  We propose a
number of axioms that such measures might be expected to satisfy, and
investigate some implications of these axioms~\footnote{We note that
Zanardi, Zalka, and Faoro~\cite{Zanardi00} pointed out the
desirability of Axioms~\ref{ax:locality} and~\ref{ax:unitary}, and of
Property~\ref{pr:exchange}, below, and proved that these properties
are all satisfied by the average entanglement generated by a
unitary.}.

%
%
The structure we adopt is to first describe (in
Sec.\,\ref{subsubse:fundamental_properties}) the fundamental axioms
that we expect {\em any} strength measure should satisfy.  We then
describe some other useful properties a strength measure may satisfy
in Sec.\,\ref{subsubse:other_properties}.  Finally,
Sec.\,\ref{subsubsec:log-rank} illustrates the axiomatic framework by
applying it to the analysis of the communication cost of distributed
quantum computation.


\subsubsection{Fundamental properties}
\label{subsubse:fundamental_properties}

%
%
We denote our strength measure by $K(\E)$, where \E\ is a
trace-preserving quantum operation acting on a set of $n$ systems,
$\A_1,\ldots,\A_n$, of dimensions $d_1,\ldots,d_n$.  We will
frequently be interested in the case where \E\ is a unitary quantum
operation $\E(\rho)=U\rho U^\dag$ for some unitary $U$.  In this case,
we write $K(U)$ to denote the dynamic strength of $U$.  We will also
use the convention that the symbol for a unitary such as $U$ may
either mean the unitary operator $U$, or the corresponding quantum
operation, that is $U(\rho)\eq U\rho U^\dag$.  This abuse of notation
will only be employed when its meaning is clear from context.

%
%
As each axiom is introduced we illustrate it by examining whether the
Hartley strength satisfies the axiom. Note that $\Khar(U)$ is defined
for a unitary operator $U$ acting on two systems labeled \A\ and \B\ 
of dimension $d_\A$ and $d_\B$, respectively.

\begin{axiom}[Non-negativity] \label{ax:non-negativity}
$K(\E)\ge0$ for all quantum operations \E.
\end{axiom}
This is more a convention than an axiom, which we introduce as a
convenience to simplify many of the properties below.  The Hartley
strength satisfies this axiom.

\begin{axiom}[Locality] \label{ax:locality}
$K(U)\ge0$ with equality if and only if $U$ can be written as a
product of local unitary operations.
\end{axiom}
The Hartley strength $\Khar(U)$ satisfies locality.

%
%
The axiom of locality captures the idea that dynamic strength measures
the {\em non-local} content of a quantum gate.  For example, in the
bipartite case, it is possible to generate entanglement with a unitary
$U$ if and only if $U$ cannot be written as a product of local unitary
operations.  Similarly, it is possible to communicate classical
information with a unitary if and only if it cannot be written as a
product of local unitaries~\cite{Beckman01}.  Summarizing,
for any $K$ satisfying locality, we have $K(U)>0$ if and only if $U$
is capable of generating entanglement or, alternatively, of
transmitting classical information.

%
%
How should the axiom of locality be extended to non-unitary
operations?  For example, we might require that $K(\E)>0$ if and only
if \E\ cannot be implemented by local operations and classical
communication.  Or perhaps we might require that $K(\E)>0$ if and only
if \E\ generates quantum states with non-zero entanglement (according
to some entanglement measure).  Many other possibilities can be
imagined which we will not enumerate.

\begin{axiom}[Local unitary invariance] \label{ax:unitary}
Suppose $A_1,\ldots,A_n$ and $B_1,\ldots,B_n$ are local unitary
operations on the respective systems $\A_1,\ldots,\A_n$.  Then
\begin{equation}
K\[(A_1\tp\cdots\tp A_n)\circ\E\circ(B_1\tp\cdots\tp
B_n)\]=K(\E).
\end{equation}
\end{axiom}
The Hartley strength satisfies local unitary invariance.

The axiom of local unitary invariance requires that the strength of a
quantum operation is not changed by local operations.  Thus, it is in
accord with the notion that the strength is a measure of an
operation's non-local content.


\subsubsection{Other useful properties}
\label{subsubse:other_properties}

%
%
We have just introduced three axioms essential for any strength
measure describing the non-local content of an operation.  We now
introduce several useful properties a strength measure may satisfy,
beginning with two invariance properties.

\begin{property}[Exchange symmetry] \label{pr:exchange}
Let \E\ be a quantum operation acting on a multipartite system whose
subsystems have the same Hilbert space.  The \swap\ operation acting
on any two of these components has the effect of interchanging their
states.  Then $K$ has the exchange symmetry property if for all such
\swap\ operations,
\begin{equation}
K(\swap\circ\E\circ\swap)=K(\E).
\end{equation}
\end{property}

\begin{property}[Time-reversal invariance] \label{pr:time}
For all unitaries $U$, $K(U^\dag)=K(U)$.
\end{property}
%
%
The Hartley strength satisfies both axioms.

\begin{property}[Continuity] \label{pr:continuity}
For some metric $D(\cdot,\cdot)$ on the space of quantum operations,
$|K(\E)-K(\F)|\le f(D(\E,\F))$, where $f(\cdot$) is a continuous and
monotone increasing function such that $f(0)=0$.
\end{property}

%
%
The Hartley strength is {\em not} continuous with respect to standard
metrics on the space of unitary operations: the presence of any
non-locality in a unitary operation $U$ is sufficient to cause a
discontinuous jump in the Hartley strength from 0 to 1 or more.

A major use of the continuity property is in the analysis of quantum
computational complexity problems; see the discussion after the
chaining property.

\begin{property}[Chaining] \label{pr:chaining}
Suppose \E\ and \F\ are two quantum operations.  Then $K(\E\circ\F)\le
K(\E)+K(\F)$.
\end{property}

%
%
The main utility of chaining was anticipated in the introduction: it
can give bounds on the number of gates required to perform a
particular quantum operation.

%
%
When combined with the continuity property, the chaining property may
also be used to prove bounds on the {\em approximation} of unitary
operations.  This is important in applications to computational
complexity since it is usually sufficient to solve problems with a
high probability of success.  Suppose, for example, that $U$ is a
desired two-qudit unitary operation, and one is given the ability to
perform a set of two-qudit gates $\U=\{U_1,\ldots,U_m\}$, and local
unitary operations.  Let $K$ be any measure satisfying continuity, for
some choice of $f$ and $D$, as above, as well as chaining and
locality.  Let $A_j\tp B_j$ be local unitaries and $U_{l_j}\in\U$.  To
obtain an approximation $V=(A_0\tp B_0)U_{l_1}(A_1\tp B_1)\cdots
U_{l_k}(A_k\tp B_k)$ to $U$ such that $D(U,V)\le\epsilon$ we need, by
the continuity property, $K(V)\ge K(U)-f(\epsilon)$.  But $K(V)\le
k\Kmax$, where \Kmax\ is the maximum value of $K(U_l)$, so the number
of gates satisfies
\begin{equation} \label{eq:prob-lower-bound}
k\ge\frac{K(U)}{\Kmax}-\frac{f(\epsilon)}{\Kmax}.
\end{equation}

%
%
The Hartley strength satisfies the chaining property, but to prove it
we need a related lemma.
\begin{lemma} \label{le:schmidt}
Suppose $U$ has operator-Schmidt decomposition $U=\sum_j s_j A_j\tp
B_j$.  Suppose $U$ can be written in some other form as a sum over
products, $U=\sum_k\tilde A_k\tp\tilde B_k$.  The number of terms in
this decomposition is at least as great as the number of terms in the
operator-Schmidt decomposition.  Thus, the operator-Schmidt
decomposition is a {\em minimal} decomposition for $U$, in the sense
that it has the fewest product terms of any sum-over-products
decomposition.
\end{lemma}

\proof A simple proof of the lemma is to note that:
\begin{equation}
A_j=\frac{1}{s_j}\tr_\B\[(I\tp B_j^\dag)U\]
=\frac{1}{s_j}\sum_k\tilde A_k\tr(B_j^\dag\tilde B_k).
\end{equation}
Thus each $A_j$ can be written as a linear combination of the $\tilde
A_k$.  But the $A_j$ are orthonormal, and thus linearly independent.
It follows that the number of operators $\tilde A_k$ must be at least
as great as the number of $A_j$, that is, at least as great as the
Schmidt number of $U$.\qed

%
%
With Lemma~\ref{le:schmidt} in hand it is straightforward to prove the
chaining property.  Suppose $U=\sum_j s_j A_j\tp B_j$ and $V=\sum_k
t_k C_k\tp D_k$ are Schmidt decompositions for unitary operators $U$
and $V$.  Then we have
\begin{equation}
UV=\sum_{jk}s_j t_k(A_j C_k)\tp(B_j D_k).
\end{equation}
The total number of terms in this sum-over-products decomposition of
$UV$ is $\Sch(U)\Sch(V)$, and so by the lemma we must have
$\Sch(UV)\le\Sch(U)\Sch(V)$.  Taking logarithms of both sides of this
inequality yields the chaining property for the Hartley strength.

%
%
Until now we have only been concerned with strength measures defined
for {\em fixed} quantum systems.  Compare this with the situation for
entanglement measures.  It is often said that there is a {\em
unique}~\cite{Popescu97,Vidal00b,Nielsen00b} entanglement measure for
bipartite pure states, namely, the von~Neumann entropy of the reduced
density matrix.  Strictly speaking, this is not a single entanglement
measure, since it can be applied to many different types of quantum
systems --- pairs of qubits, a qubit and a qutrit, and so on.  Rather,
it is a {\em family} of entanglement measures, satisfying certain
consistency properties that make it sensible to refer to it as a
single measure.

%
%
Motivated by this, we describe two consistency properties we expect of
a family of strength measures.  There are two different ways in which
a family of strength measures arises naturally.  The first corresponds
to appending additional systems while keeping the state-space
dimensions of the existing systems constant.  The second corresponds
to fixing the number of systems, and varying the state-space
dimensions of the individual systems by adding local
ancillas.

%
%
For the statement of each of the following properties we imagine that
there is a {\em family} of strength measures, each of which is denoted
by the same letter $K$.  When necessary, we add superscripts to make
precise which systems $K$ is acting on.  For example,
$K^{\A:\B:\C}(\E)$ indicates the strength with respect to a division
into three components, labeled \A, \B, and \C, and $K^{\A:\B\C}(\E)$
indicates the strength with respect to a division into two components
\A\ and \B\C.  For notational simplicity, we state these properties
for the case of three systems, with the generalization to more systems
following similar lines.
\begin{property}[Stability under addition of systems]
\label{pr:systems}
Suppose \E\ acts on systems \A\ and \B, and \C\ is an additional
system.  Then the family $K$ is stable with respect to additional
systems if
\begin{equation}
K^{\A:\B}(\E)\ge K^{\A:\B:\C}(\E\tp\I),
\end{equation}
where \I\ denotes the identity operation on \C.
\end{property}
%
%
Note that it does not make sense to speak of the Hartley strength as
being stable or not stable in this sense, since it is only defined for
two-component systems.

%
%
The intuition motivating the inequality in the statement of stability
is that the ``two-party'' non-locality present in \E\ should not be
less than the ``three-party'' non-locality in $\E\tp\I$. A stronger
statement of the stability property would replace the inequality by an
equality.

%
%
The stability property is useful in the context of quantum
computational complexity.  We explained earlier how to derive lower
bounds such as Eq.\,(\ref{eq:lower-bound}) and
Eq.\,(\ref{eq:prob-lower-bound}) on the number of gates needed to
implement a two-qudit quantum operation.  In the context of quantum
computational complexity, the most natural setting is that we wish to
implement a family of $n$-qubit unitaries $U$ (indexed by $n$) using a
universal set of one- and two-qubit quantum gates.  In such a setting,
we are looking for the most efficient decomposition of each $U$ into a
product of two-qubit gates
\begin{equation}
U=U_{j_1 k_1}U_{j_2 k_2}\cdots U_{j_l k_l},
\end{equation}
where the subscripts denote the qubits on which each (possibly
different) unitary gate acts.  A bound on the minimum number of gates
$l$ may be deduced from the chaining and stability properties, using a
similar analysis to that given in connection with chaining alone,
$l\ge K(U)/\Kmax$, where now \Kmax\ is the maximum value of the
strength of any two-qubit gate.  Because of stability, \Kmax\ is a
{\em constant}, independent of $n$, so in order to prove interesting
lower bounds on $l$, one only needs to analyze the asymptotic behavior
of $K(U)$ as a function of $n$.  If, for example, we could find a
strength measure satisfying both chaining and stability, and such that
$K(U)=\Theta(2^n)$ for some family of unitaries $U$, then it would
follow that the family requires a number of gates exponential in $n$.
If, in addition, $K$ has suitable continuity properties, then it may
be possible to prove that the family requires exponential time even if
some reasonable probability of error is allowed.  Needless to say, if
this were true for a unitary encoding of, say, the solution to a
problem such as the traveling salesman problem, this would be a very
interesting result indeed.

%
%
Our second notion of stability is that introducing local ancillas
which are then ignored should not change the strength of an operation.

\begin{property}[Stability with respect to ancillas]
\label{pr:ancilla}
Suppose \E\ acts on systems \A\ and \B, and \C\ is an additional
system.  Then the family $K$ is stable with respect to local ancillas
if
\begin{equation}
K^{\A:\B}(\E)=K^{\A:\B\C}(\E\tp\I).
\end{equation}
\end{property}
%
%
The Hartley strength is clearly stable with respect to local ancillas.

%
%
We now move on to additivity properties.

\begin{property}[Weak (sub)additivity] \label{pr:weak}
Suppose $\A_1$, $\A_2$, $\B_1$, and $\B_2$ are distinct systems such
that $\A_1$ and $\A_2$ have the same state space, as do $\B_1$ and
$\B_2$.  Suppose \E\ is a quantum operation that can act on either
$\A_1\B_1$ or $\A_2\B_2$.  Then the family $K$ is {\em weakly
subadditive} if
\begin{equation}
K^{\A_1\A_2:\B_1\B_2}(\E\tp\E)\le2K^{\A_1:\B_1}(\E).
\end{equation}
$K$ is {\em weakly additive} if the inequality can be replaced by an
equality in the above expression.
\end{property}

\begin{property}[Strong (sub)additivity] \label{pr:strong}
Suppose $\A_1$, $\A_2$, $\B_1$, and $\B_2$ four distinct systems, and
\E\ and \F\ are quantum operations acting on $\A_1\B_1$ and
$\A_2\B_2$, respectively.  Then the family $K$ is {\em strongly
subadditive} if
\begin{equation}
K^{\A_1\A_2:\B_1\B_2}(\E\tp\F)\le K^{\A_1:\B_1}(\E)+K^{\A_2:\B_2}(\F).
\end{equation}
$K$ is {\em strongly additive} if the inequality can be replaced by an
equality in the above expression.
\end{property}
%
%
Note that strong subadditivity for a strength measure is not connected
with the strong subadditivity property for quantum mechanical
entropy~\cite{Lieb73b}.

%
%
The Hartley strength is strongly additive for unitary operations $U$
and $V$, and thus possesses all four of these properties.  To see
this, suppose $U$ and $V$ are unitary operators with Schmidt
decompositions $U=\sum_j s_j A_j\tp B_j$ and $V=\sum_k t_k C_k\tp
D_k$, where $A_j$, $B_j$, $C_k$, and $D_k$ act on systems $\A_1$,
$\B_1$, $\A_2$,and $\B_2$, respectively.  Then the Schmidt
decomposition of $U\tp V$ with respect to $\A_1\B_1:\A_2\B_2$ is
\begin{equation} \label{eq:tp-Schmidt}
U\tp V=\sum_{jk}s_j t_k(A_j\tp C_k)\tp(B_j\tp D_k).
\end{equation}
It follows that $\Sch(U\tp V)=\Sch(U)\Sch(V)$ and, taking logarithms,
we see that the Hartley strength is strongly additive.

\begin{proposition} \label{po:subadditive}
If the family $K$ satisfies the chaining property and is stable
with respect to local ancillas, then it is strongly subadditive.
\end{proposition}

\proof Applying simple algebra, the chaining property, and stability
with respect to local ancillas in turn, we have:
\begin{equation} \begin{split}
K(\E\tp\F)&=K\[(\E\tp\I)(\I\tp\F)\]\\
&\le K(\E\tp\I)+K(\I\tp\F)\\
&\le K(\E)+K(\F),
\end{split} \end{equation}
which is the strong subadditivity property.\qed

The converse is not true --- we will see later that the Schmidt
strength is strongly additive and stable with respect to local
ancillas, but does not satisfy chaining.

%
%
The final property addresses what happens when a quantum operation
arises as a consequence of tracing out part of the action of a quantum
operation acting on a larger system.  For notational simplicity, we
state this property for the special case of two systems, with the
generalization to more systems following similar lines.

\begin{property}[Reduction] \label{pr:reduction}
Suppose a quantum operation \E\ on a composite system \A\B\ is
obtained from a quantum operation on \A\B\C\ as follows:
\begin{equation}
\E(\rho_{\A\B})=\tr_\C\[\F(\rho_{\A\B}\tp\sigma_\C)\],
\end{equation}
for some fixed state $\sigma_\C$ of system \C.  Then a family $K$ of
strength measures has the {\em reduction property} if
$K^{\A:\B}(\E)\le K^{\A:\B\C}(\F)$.
\end{property}
%
%
The intuition behind the reduction property is that if it is possible
to do \F, then it is also possible to do \E, without any extra
dynamical resources being required.  

%
%
The reduction property is important both in the analysis of
distributed quantum computation (see below) and for the applications
to quantum computational complexity suggested earlier in this paper.
In the latter applications we implicitly assumed that the
implementation of some desired unitary could not be assisted by the
introduction of ancilla qubits that are discarded at the end of the
computation.  However, there is evidence to suggest that ancilla may
help in performing a unitary transformation quickly; for example, some
of the constructions in~\cite{Barenco95} were made more efficient by
the use of ancilla.  Suppose, however, that $K$ has the reduction
property, and that $U$ can be implemented by performing an operation
$V$ on a larger system.  That is, suppose
$V|\psi\>|s\>=(U|\psi\>)|s'\>$, for all $|\psi\>$, and for some fixed
ancilla states $|s\>$ and $|s'\>$.  Then we have $K(U)\le K(V)$.  If,
in addition, it is possible to use $K(\cdot)$ to prove bounds on
computational complexity, as described earlier, then it follows from
the inequality $K(U)\le K(V)$ that any bound on the computational
complexity of $U$ must also apply to $V$, and thus our techniques can
be applied even when working qubits are allowed.

%
%
The reduction property makes restricted sense for the Hartley
strength, which is defined only for unitary operators.  In particular,
imagine, as above, that we have a unitary $V$ acting on \A\B\C\ such
that $V|\psi\>|s\>=(U|\psi\>)|s'\>$, where $|\psi\>$ is an arbitrary
state of \A\B, $U$ is a unitary acting on \A\B\ alone, and $|s\>$ and
$|s'\>$ are fixed states of \C.  To see that \Khar\ satisfies the
reduction property, let us introduce orthonormal bases $|j\>$, $|k\>$,
and $|l\>$ for the systems \A, \B, and \C, respectively.  Note that
the invariance of \Khar\ with respect to unitaries on system \C\
implies that it suffices to consider $V$ such that
$V|\psi\>|0\>=U|\psi\>|0\>$, where $|0\>$ is the first element of the
basis for $\C$.  Suppose we expand $V$ as
\begin{equation}
V=\sum_{jklj'k'l'}V_{j'k'l',jkl}|j'\>\<j|\tp|k'l'\>\<kl|,
\end{equation}
where the comma in the subscript of $V$ separates the row index from
the column index.  Since $|j'\ket \bra j|$ and $|k'l'\ket \bra kl|$
are orthonormal operator bases, it follows that the Schmidt
coefficients of $V$ are just the singular values of the matrix $\tilde
V$ defined by $\tilde V_{jj',kk'll'}\eq V_{j'k'l',jkl}$.  Thus, the
Schmidt number of $V$ is given by the number of non-zero singular
values, or the {\em rank}, of the matrix .  Similarly, we can expand
$U$ as
\begin{equation}
U=\sum_{jkj'k'}U_{j'k',jk}|j'\>\<j|\tp|k'\>\<k|,
\end{equation}
and the Schmidt number of $U$ is given by the rank of the matrix
$\tilde U_{jj',kk'}\eq U_{j'k',jk}$.  But $U_{j'k',jk}=V_{j'k'0,jk0}$,
so up to reordering of the columns $\tilde V=[\tilde U|\cdots]$.  It
follows that the rank of $\tilde V$ is at least as great as the rank
of $\tilde U$, and thus $\Sch(V)\ge\Sch(U)$.  Taking logarithms of
both sides we get $\Khar(V)\ge\Khar(U)$, which is the reduction
property.


\subsubsection{Application to the log-rank lower bound}
\label{subsubsec:log-rank}

%
%
As an illustration of the power of the framework we have just
developed, we now apply it to the analysis of a computational problem
of considerable interest: the communication cost of a distributed
computation.

%
%
We consider two separate problems in distributed computation, the
first related to distributed computation of a classical function, the
second to distributed computation of the quantum Fourier transform.
The first problem may be stated as follows.  Suppose Alice and Bob are
in possession of classical data strings $x$ and $y$, respectively.
They wish to compute some joint one-bit function $f(x,y)$ of the data
strings.  To accomplish this task they are only able to do arbitrary
local quantum operations and to communicate qubits.  This is the key
problem of {\em quantum communication
complexity}~\cite{Yao93,Kremer95}.

%
%
One of the major results in the fields of quantum and classical
communication complexity is the {\em log-rank lower bound}.  This
states that the minimum number of bits (or qubits) of communication
required to compute $f(x,y)$ is bounded below by
$\log[\text{rank}(-1^{f(x,y)})]$, where $(-1)^{f(x,y)}$ is the
$(x,y)$th entry of the {\em communication matrix}.  Mehlhorn and
Schmidt~\cite{Mehlhorn82} proved this result for classical
communication complexity.  The {\em log-rank conjecture} of
communication complexity~\cite{Kushilevitz97} states that, up to a
polynomial factor, the log-rank lower bound is {\em saturated}, that
is, there is a protocol to compute $f(x,y)$ using
$\text{polylog}[\text{rank}(-1^{f(x,y)})]$ bits of communication.

%
%
Although quantum protocols are potentially more powerful than
classical, it was pointed out by Buhrman, Cleve and
Wigderson~\cite{Buhrman98} that \cite{Yao93,Kremer95} contains an
implicit proof of the log-rank lower bound in the quantum case.  This
result was extended to the model in which pre-shared entanglement is
allowed by Buhrman and de Wolf~\cite{Buhrman99}.

%
%
The framework introduced above and the results we have proved about
the Hartley strength allow us to give an almost trivial proof of the
log-rank lower bound in the case when only qubit communication is
allowed, with no pre-shared entanglement.  The proof is as follows.
Suppose we have a protocol in which Alice and Bob compute $f(x,y)$
using $k$ qubits of communication.  Then it is not difficult to see
that they can also compute $f(x,y)$ using at most $k$ \swap\ gates and
no qubit communication.  Using Bennett's techniques~\cite{Bennett73}
of reversible computation, the protocol may be modified (using only
local unitary operations) to give what Cleve \etal~\cite{Cleve97b}
called a {\em clean protocol} effecting the unitary transformation
$|w_A\>|x\>|y\>|w_B\>\to(-1)^{f(x,y)}|w_A\>|x\>|y\>|w_B\> $, where
$|w_A\>$ and $|w_B\>$ are local work qubits for Alice and Bob.  The
clean protocol uses only $2k$ \swap\ gates.  Let $V$ be the unitary
effected by the clean protocol, and let
$U|x\>|y\>\eq(-1)^{f(x,y)}|x,y\>$.  Then by the reduction property
followed by the chaining property we have
\begin{equation} \label{eq:log-rank-inter}
\Khar(U)\le\Khar(V)\le2k\Khar(\swap)=4k.
\end{equation}
But $U=\sum_{xy}(-1)^{f(x,y)}|x\>\< x|\tp|y\>\< y|$ from which
it follows that $\Sch(U)=\text{rank}((-1)^{f(x,y)})$.  Combining this
observation with Eq.\,(\ref{eq:log-rank-inter}) gives the log-rank
lower bound
\begin{equation}
k\ge\frac{1}{4}\log\[\text{rank}\(-1^{f(x,y)}\)\].
\end{equation}

%
%
The second problem in distributed computation we consider is the
distributed computation of a unitary operation such as the quantum
Fourier transform $U$ on $m+n$ qubits ($m \leq n$), where Alice is in
possession of the first $m$ qubits, and Bob is in possession of the
remaining $n$ qubits~\footnote{The following discussion generalizes
results in~\cite{Nielsen98}, which considered the case $m=n$.}.  How
many qubits of communication must Alice and Bob do to compute $U$?
Suppose it is possible to achieve it with just $k$ qubits of
communication.  Then, as in the discussion of the log-rank lower
bound, it must also be possible to implement the quantum Fourier
transform in a model in which no qubit communication is allowed, but
in which Alice and Bob can apply $k$ \swap\ gates to their qubits.
Applying the reduction and chaining properties we conclude that
$\Khar(U)\le k\Khar(\swap)$, and thus we obtain the lower bound $k\ge
2m$, which agrees with the $m=n$ result obtained in~\cite{Nielsen98}.


\section{The canonical decomposition} \label{se:interlude}

Before we describe our results about measures of dynamic strength, we
pause to explore a useful representation theorem for two-qubit unitary
operators, the {\em canonical decomposition} of Khaneja, Brockett, and
Glaser~\cite{Khaneja01} (see also Kraus and Cirac~\cite{Kraus01} for a
simple, constructive proof).  This decomposition is an extremely
valuable tool which characterizes the non-local properties of any
two-qubit unitary with only three parameters, $\theta_x$, $\theta_y$,
and $\theta_z$~\footnote{See Makhlin~\cite{Makhlin00} for an
earlier proof that the non-local properties of $U$ are
characterized by $\theta_x$, $\theta_y$ and $\theta_z$.}.  For
appropriate one-qubit unitaries $A_1$, $A_2$, $B_1$, and $B_2$,
\begin{equation}
U=(A_1\tp B_1)e^{i(\theta_x X\tp X+\theta_y Y\tp Y+\theta_z Z\tp
Z)}(A_2\tp B_2)
\end{equation}
where $-\frac{\pi}{4}<\theta_\alpha\le\frac{\pi}{4}$.  For
convenience, define the {\em canonical form} of $U$ to be $\tilde
U\eq(A_1^\dag\tp B_1^\dag)U(A_2^\dag\tp B_2^\dag)$; up to local
unitaries, $\tilde U$ is equivalent to $U$.

Since $X\tp X$, $Y\tp Y$, and $Z\tp Z$ all commute, we may expand
$\tilde U$ as
\begin{eqnarray}
\tilde U&\!\!=&\!\!\(c_x I\tp I+is_x X\tp X\)\times\(c_y I\tp I+i
s_y Y\tp Y\)\non\\
&&\times\(c_z I\tp I+is_z Z\tp Z\),
\end{eqnarray}
where $c_\alpha\eq\cos(\theta_\alpha)$,
$s_\alpha\eq\sin(\theta_\alpha)$.  Multiplying the expression out
yields
\begin{equation} \begin{split} \label{eq:two-qubit-special-form}
\tilde U=(c_x c_y c_z+is_x s_y s_z)I\tp I+(c_x s_y s_z+is_x c_y
c_z)X\tp X\\
+\:(s_x c_y s_z+ic_x s_y c_z)Y\tp Y+(s_x s_y c_z+ic_x c_y s_z)Z\tp Z.
\end{split} \end{equation}
This expression is essentially in Schmidt form: up to a constant the
Schmidt coefficients are just the magnitudes of the coefficients
appearing in front of the four terms.
Eq.\,(\ref{eq:two-qubit-special-form}) enables us to deduce the
following result~\footnote{After completion of this work, we learnt
  that an equivalent result for states has been independently obtained
  by D\"ur, Vidal and Cirac~\cite{Dur02}.}:
\begin{proposition}
There exist two-qubit unitary operators with Schmidt number $1$, $2$,
and $4$, but not $3$.
\end{proposition}
This is a surprising result because it reveals unexpected structure in
the space of two-qubit unitary operators.  It is tempting to speculate
on the existence of similar structure for more general unitary
operators.  We conjecture that, in a $d\times d'$ system, there exist
unitary operators with Schmidt number $k$ if and only if $k$ divides
$dd'$.  An alternative conjecture, which we believe is less likely, is
that unitary operators with Schmidt number $k$ exist if and only if
$k$ and $dd'$ are not coprime.

\proof It is straightforward to see that unitaries with Schmidt number
1, 2, and 4 exist, so it only remains to show that there exist none
with Schmidt number 3.  Suppose $U$ has Schmidt number 3.  Then the
canonical form of $U$, $\tilde U$, must have exactly one of the terms
in Eq.\,(\ref{eq:two-qubit-special-form}) equal to zero.  Without loss
of generality, suppose the $I\tp I$ term is zero.  (If, for example,
the $X \tp X$ term is zero, then we multiply by $X\tp X$ to obtain a
unitary with the $I\tp I$ term zero.)  Then we must have $c_x c_y
c_z=s_x s_y s_z=0$, and therefore $c_\alpha=0$ for at least one value
of $\alpha$, and $s_\beta=0$ for at least one value of $\beta$.  Note
that $\alpha$ cannot be equal to $\beta$ since
$s_\alpha^2+c_\alpha^2=1$.  By symmetry it suffices to assume that
$\alpha$ is $x$ and $\beta$ is $y$, in which case we obtain a unitary
of the form $is_x c_y c_z X\tp X+s_x c_y s_z Y\tp Y$, which has
Schmidt number at most 2.\qed

%
%
Now suppose that $U$ has $\Sch(U)\le2$.  Then, up to local unitary
operations, it has the form of Eq.\,(\ref{eq:two-qubit-special-form}),
with exactly two of the terms non-zero.  As mentioned in the previous
proof, we can always ensure that the $I\tp I$ term is non-zero.
Furthermore, conjugating by local unitaries, we can ensure that the
other non-zero term is $X\tp X$.  Thus, up to local unitary
equivalence, $U$ has the form $U=aI\tp I+bX\tp X$, for some non-zero
$a$ and $b$.  Furthermore, we may assume that $a$ is real, since we
can multiply $U$ by the local unitary operation $(e^{i\phi}I)\tp I$.
Unitarity of $U$ then implies that
\begin{equation}
I\tp I=U^\dag U=(a^2+|b|^2)I\tp I+a(b^*+b)X\tp X,
\end{equation}
from which we deduce that $a^2+|b|^2=1$ and $a(b^*+b)=0$.  Since
$a\ne0$, $b$ must be pure imaginary.  Thus we have $a=\sqrt{1-p}$,
$b=i\sqrt p$ for some $0\le p\le1$.  We have proved the following:
\begin{proposition} \label{po:schmidt2}
Let $U$ be a two-qubit unitary operator with Schmidt number $2$.
Then, up to local unitary equivalence, $U$ has the form:
\begin{equation} \label{eq:controlled-X}
U=\sqrt{1-p}\,I\tp I+i\sqrt{p}\,X\tp X.
\end{equation}
\end{proposition}


\section{Strength measures based on entanglement generation}
\label{se:generation}

In this and the following section we explore some of the strength
measures defined by us and other authors, noting relations between
them, and connections to our earlier operational questions.  We also
prove several results about which measures obey which
axioms/properties, summarized in Table~\ref{ta:summary} at the end of
this paper.

We start in this section with strength measures based on entanglement
generation.  More is known about these measures because they use the
relatively well-developed field of state entanglement.  It seems
likely to us that, although these are natural measures to consider
first, in the long run they may not be the most useful.  Since they
are based on static resources, they may not provide much more insight
when applied to dynamics.  We consider two classes of
entanglement-based strength measures: the entanglement generating
capacities of quantum operations \emph{without} initial entanglement,
and entanglement generating capacities \emph{with} the possibility of
initial entanglement.


\subsection{Entanglement generation without prior
  entanglement}

Recall the definition of $\Ke(U)$, Eq.\,(\ref{eq:kanc}):
$\Ke(U)=\max_{|\alpha\>,|\beta\>}E(U|\alpha\>|\beta\>)$.  $\Ke(U)$
measures the maximum amount of entanglement generated by a single
application of the unitary $U$ without initial entanglement.  We show
that \Ke\ and \Ksch\ are related to each other in interesting ways:
(1) \Ksch\ is a lower bound for \Ke; and (2) \Ke\ is equal to \Ksch\ 
for a class of two-qubit unitaries.  We also give some numerical
evidence demonstrating that \Ke\ is not equal to \Ksch\ for certain
unitaries; see Fig.~\ref{fi:kanc-neq-ksch}.  To make this discussion
easier, we begin by discussing of the properties satisfied by \Ke\ and
\Ksch, including a demonstration of the striking property that \Ke\ is
{\em superadditive}, that is $U\tp U$ can sometimes generate strictly
more than twice as much entanglement as $U$ alone.  Finally, we extend
the definition of \Ke\ and \Ksch\ to general quantum operations, and
prove that $\Ke\ge\Ksch$ still holds.


\subsubsection{Properties of \Ke\ and \Ksch}

Beginning with the three axioms, it is easy to see that both \Ke\ and
\Ksch\ satisfy non-negativity, locality, and local unitary invariance.
(As we have only defined \Ke\ and \Ksch\ 
for unitaries, the axioms and properties we discuss here are
restricted to this case.) 

We now turn to the properties of \Ksch, which are very similar to
those of \Khar.  \Ksch\ clearly satisfies the properties of exchange
symmetry, time-reversal invariance, and stability with respect to
local ancillas, since none of these operations change the Schmidt
coefficients.  The argument that \Ksch\ is continuous is slightly
complicated, and will be described in the next paragraph.  \Ksch\ is
strongly additive, \ie $\Ksch(U\tp V)=\Ksch(U)+\Ksch(V)$.  To see
this, recall that if $U$ and $V$ have Schmidt decompositions $U=\sum_j
s_j A_j\tp B_j$ and $V=\sum_k t_k C_k\tp D_k$, with $A_j$, $B_j$,
$C_k$ and $D_k$ acting on systems $\A_1$, $\A_2$, $\B_1$, and $\B_2$,
respectively, then the Schmidt decomposition of $U\tp V$ with respect
to $\A_1\B_1:\A_2\B_2$ is given by Eq.\,(\ref{eq:tp-Schmidt}):
\begin{equation}
U\tp V=\sum_{jk}s_j t_k(A_j\tp C_k)\tp(B_j\tp D_k)\non.
\end{equation}
Using properties of the Shannon entropy, we find that
\begin{equation} \begin{split}
\Ksch(U\tp V)&=H\textstyle{\(\l\{\frac{s_l^2 t_k^2}{d_\A^2
d_\B^2}\r\}\)}\\
&=H\textstyle{\(\l\{\frac{s_l^2}{d_\A
d_\B}\r\}\)}+H\textstyle{\(\l\{\frac{t_k^2}{d_\A d_\B}\r\}\)}\\
&=\Ksch(U)+\Ksch(V).
\end{split} \end{equation}

%
%
To see that \Ksch\ is continuous, expand 
\begin{eqnarray}
U = \sum_{jj'kk'} U_{jj',kk'} |j\ket \bra k| \tp |j'\ket \bra k'|,
\end{eqnarray}
where the comma separates row and column indices.  Since $|j\ket \bra
k|$ and $|j'\ket \bra k'|$ are orthonormal operator bases, it follows
that the Schmidt coefficients of $U$ are just the singular values of
the matrix $\tilde U$ defined by $\tilde U_{jk,j'k'} \equiv
U_{jj',kk'}$.  Consider the matrix norm $\| A \| \equiv
\max_{|\psi\ket} \| A |\psi\ket \|$, where the maximization is over
unit vectors $|\psi\ket$.  \Ksch\ is a continuous function of the
Schmidt coefficients, and the Schmidt coefficients are continuous
functions of the matrix $U$, with respect to matrix norm.  This
follows from the fact that the singular values of a matrix are
continuous in the matrix (see, e.g., Chapter~3 of~\cite{Horn91}).
Thus \Ksch\ is a continuous function of $U$ with respect to the matrix
norm.

\begin{center}
\begin{figure}[ht]
\scalebox{0.6}{\includegraphics{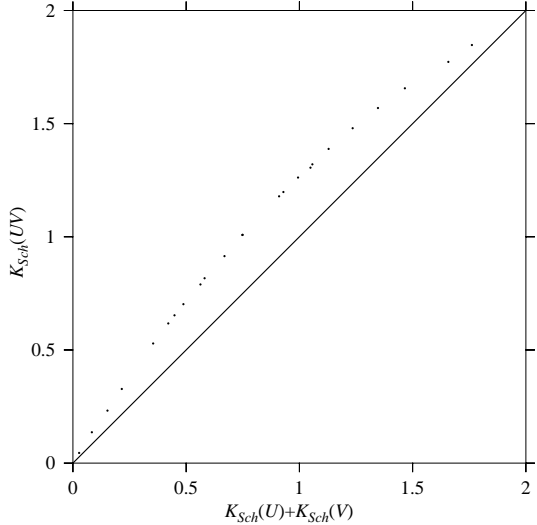}}
\caption{\label{fi:chaining}Numerical violation of the chaining
property for \Ksch.  $U$ and $V$ are two-qubit unitaries chosen by
first generating random unitaries, and then using a Nelder-Mead
simplex minimization algorithm to prepend and append local unitaries
to generate $U$ and $V$ maximizing the violation of $\Ksch(UV) \leq
\Ksch(U)+\Ksch(V)$.  If \Ksch\ satisfied chaining, then all the points
(corresponding to pairs of unitaries $U$ and $V$) would lie on or
below the line.}
\end{figure}
\end{center}

We have demonstrated numerically that \Ksch\ does not satisfy
chaining; see Fig.\,\ref{fi:chaining}.
%
%
\Ksch\ also violates the reduction property.  To see this, suppose a
Toffoli gate $V$ is applied to three qubits \A\B\C, with \A\ acting as
the target qubit.  Suppose \C\ is initially prepared in the $|1\>$
state, so $V|\psi\>|1\>=(U|\psi\>)|1\>$, where $U$ is the \cnot\ gate,
and $|\psi\>$ is an arbitrary state of \A\B.  It is not difficult to
verify that $\Ksch(U)=H(1/2,1/2)$, while $\Ksch(V)=H(1/4,3/4)$, so
$\Ksch(V)<\Ksch(U)$, in violation of the reduction property.

%
%
The properties of \Ke\ are somewhat more difficult to elicit.  \Ke\ is
easily seen to satisfy the exchange symmetry property.  Numerical
studies of the time-reversal invariance property have been
inconclusive, although we speculate that for two-qutrit unitaries
time-reversal invariance will \emph{not} be obeyed.  The discussion of
continuity is somewhat complicated, and is described in the following
paragraph.  \Ke\ is stable with respect to ancillas, since it already
allows for the possibility of arbitrary ancillas.  It is also easy to
see from the definition that \Ke\ satisfies the reduction property, in
the same sense that the Hartley strength satisfies the reduction
property.

%
%
We now outline a proof that \Ke\ is continuous.  To prove this, we
need to introduce a metric on the space of unitary matrices.  We use
the matrix norm, $D(U,V) \equiv \| U-V\| = \max_{|\psi\ket} \|
(U-V)|\psi\ket \|$, where the maximum is over all unit vectors
$|\psi\ket$.  Choose $|\alpha\ket,|\beta\ket$ such that $\Ke(V) =
E(V|\alpha\ket|\beta\ket)$.  Our earlier discussion shows that,
without loss of generality, we may assume $|\alpha\ket$ lives in a
$d_{\A}^2$-dimensional space, and $|\beta\ket$ lives in a
$d_{\B}^2$-dimensional space.  It follows from the definition that
\begin{eqnarray} \label{eq:Ke-bound}
\Ke(U) \geq E(U|\alpha\ket|\beta\ket)
\end{eqnarray}
The results of~\cite{Nielsen00b} (see also~\cite{Donald99}) imply
that, provided $\|\,|\psi\ket-|\phi\ket\|\le1/6$,
\begin{eqnarray}
|E(|\psi\ket)-E(|\phi\ket)|&\!\!\le&\!\!2\|\,|\psi\ket-|\phi\ket \|
\log(d_\C d_\D) \non\\
&&+\,\eta(2\|\,|\psi\ket-|\phi\ket\|),
\end{eqnarray}
where $\eta(x)=-x\log(x)$.  Thus, provided $\|U-V\|\le1/6$,
\begin{eqnarray}
|E(U|\alpha\ket|\beta\ket)-E(V|\alpha\ket |\beta\ket)| 
&\!\!\le&\!\!2\|U-V\|\log(d_\A^2d_\B^2) \nonumber \\
&&+\,\eta(2\| U-V\|).
\end{eqnarray}
Combining this result with Eq.\,(\ref{eq:Ke-bound}) and the fact that
$\Ke(V) = E(V|\alpha\ket |\beta\ket)$, we obtain
\begin{equation}
\Ke(U)\geq
\Ke(V) -2\| U-V\| \log(d_{\A}^2d_{\B}^2) -\eta(2\| U-V\|).
\end{equation}
By symmetry the same inequality holds with $U$ and $V$ interchanged,
and thus
\begin{equation}
|\Ke(U)-\Ke(V)|\le4\|U-V\|\log(d_{\A}d_{\B})+\eta(2\| U-V\|)
\end{equation}
whenever $\|U-V\| \leq 1/6$, which is the desired continuity equation.

%
%
What about the additivity properties of \Ke?  Intuitively, we expect
the amount of entanglement generated by two copies of $U$ is no
greater than twice the maximum generated by one use of $U$.  However,
this intuition fails when ancillas are allowed.  We show below that,
unlike \Ksch, \Ke\ is superadditive.  The proof requires some facts
about the relationship between \Ke\ and \Ksch, so we prove this result
at the end of Sec.\,\ref{subsubse:relations}.  Since \Ke\ is stable
with respect to local ancillas, subadditivity of \Ke\ and
Proposition\,\ref{po:subadditive} imply that \Ke\ does not satisfy
chaining.


\subsubsection{Relations between \Ke\ and \Ksch}
\label{subsubse:relations}

In this subsection, we explore some relations between \Ksch\ and \Ke.

%
%
\begin{lemma} \label{le:ksch}
For all unitaries $U$, $\Ksch(U)=E(U|\alpha\>|\beta\>)$ where
$|\alpha\>$ is a maximally entangled state of system \A\ with an
ancilla $\R_\A$, and $|\beta\>$ is a maximally entangled state of
system \B\ with an ancilla $\R_\B$.
\end{lemma}
\proof Let \A\ and \B\ label Alice's and Bob's systems, respectively.
Alice introduces an ancilla $\R_\A$ that is a copy of her system.  She
prepares \A\ and $\R_\A$ in a maximally entangled state,
$|\alpha\>=\frac{1}{\sqrt{d_\A}}\sum\nolimits_j|j\>|j\>$, where $d_\A$
is the dimension of system \A\ (and hence also of system $\R_\A$).
Bob does the same thing, preparing
$|\beta\>=\frac{1}{\sqrt{d_\B}}\sum\nolimits_j|j\>|j\>$, where $d_\B$
is similarly the dimension of \B.

Let $U=\sum_l s_l A_l\tp B_l$ be the Schmidt decomposition of $U$
(Eq.\,(\ref{eq:operator_schmidt})).  Alice and Bob apply $U$ to \A\B,
obtaining
\begin{equation}
U|\alpha\>|\beta\>=\sum_l s_l A_l|\alpha\>B_l|\beta\>=\sum_l
\frac{s_l}{\sqrt{d_\A d_\B}}|a_l\>|b_l\>,
\end{equation}
where we define $|a_l\>\eq\sqrt{d_\A}A_l|\alpha\>$ and
$|b_l\>\eq\sqrt{d_\B}B_l|\beta\>$.  $|a_l\>$ and $|b_l\>$ are
orthonormal bases.  For example:
\begin{equation}
\<a_k|a_l\>=d_\A\<\alpha|A_k^\dag A_l|\alpha\>=\tr A_k^\dag A_l=
\delta_{kl}.
\end{equation}
Therefore, $U|\alpha\>|\beta\>$ has entanglement $H\(\l\{s_l^2/(d_\A
d_\B)\r\}\)$ which is equal to $\Ksch(U)$.\qed

%
%
{}From this lemma, it follows that $\Ke(U)$ is bounded below by
$\Ksch(U)$.  We also show that they are equal for certain two-qubit
unitaries:

\begin{theorem} \label{th:kanc-geq-ksch}
$\Ke(U)\ge\Ksch(U)$ for all unitaries $U$.
\end{theorem}

\begin{theorem} \label{th:kanc-eq-ksch}
$\Ke(U)=\Ksch(U)$ for all two-qubit unitaries $U$ satisfying
$\Sch(U)\le2$.
\end{theorem}
\proof When $\Sch(U)=1$, $U$ is a local unitary and hence
$\Ke(U)=\Ksch(U)=0$.

Suppose $\Sch(U)=2$, in which case Proposition~\ref{po:schmidt2}
implies that $U$ may be expanded as
\begin{equation}
U=(A_1\tp B_1)\(\sqrt{1-p}\,I\tp I+i\sqrt{p}\,X\tp X\)(A_2\tp B_2).
\end{equation}
Let $\tilde U=\sqrt{1-p}\,I\tp I+i\sqrt{p}\,X\tp X$.  We have seen in
the previous section that \Ke\ and \Ksch\ are both invariant under
local unitaries, so we have $\Ke(U)=\Ke(\tilde U)$ and
$\Ksch(U)=\Ksch(\tilde U)$.

We can calculate $\Ksch(\tilde U)$ and $\Ke(\tilde U)$ directly.
$\Ksch(\tilde U)$ is equal to $H(1-p,p)\eq H(p)$, the binary Shannon
entropy.  To calculate $\Ke(\tilde U)$, we substitute $\tilde U$ into
the expression Eq.\,(\ref{eq:kanc}) for $\Ke$, giving
\begin{equation} \begin{split}
\Ke(\tilde U)&=\max_{|\alpha\>,|\beta\>}S\Big[(1-p)|\alpha\>\<\alpha|+
pX|\alpha\>\<\alpha|X\\
+\,i&\sqrt{p(1-p)}\<\beta|X|\beta\>(X|\alpha\>\<\alpha|-|\alpha\>\<
\alpha|X)\Big],
\end{split} \end{equation}
where $S$ is the von Neumann entropy, and its argument 
is a state of $\A\R_\A$.
Now we use the fact that a projective measurement on
$\A\R_\A$ cannot decrease its entropy (see Chapter~11
of~\cite{Nielsen00}).  We measure in an orthonormal basis containing
the elements $|\alpha\>$ and $|\alpha_\perp\>$, where
$|\alpha_\perp\>$ is chosen so that, up to an unimportant global
phase, $X|\alpha\>=\cos\phi|\alpha\>+\sin\phi|\alpha_\perp\>$ for some
$\phi$.  We obtain
\begin{eqnarray} \label{eq:measure}
\Ke(\tilde U)&&\!\!\!\!\!\!\le\max_{|\alpha\>}S\[(1-p)|\alpha\>\<\alpha|+p\<
\alpha|X|\alpha\>^2|\alpha\>\<\alpha|\r.\non\\
&&\ \ \ \ \ \ \ \ \ \ \ \l.+\,p|\<\alpha_\perp|X|\alpha\>|^2|
\alpha_\perp\>\<\alpha_\perp|\]\non\\
&&\!\!\!\!\!\!\le\max_\phi H(1-p+p\cos^2\phi,p\sin^2\phi).
\end{eqnarray}
If $p\le\frac{1}{2}$, the maximum occurs for $\phi=\frac{\pi}{2}$ and
$\Ke(U)\le H(p)=\Ksch(U)$.  (If $p>\frac{1}{2}$, apply $X\tp X$
to $U$ to swap the role of $p$ and $1-p$.)  Since, by
Th.\,\ref{th:kanc-geq-ksch}, $\Ke(U)$ is greater than or equal
to $\Ksch(U)$, we must have equality.\qed

\begin{center}
\begin{figure}[ht]
\scalebox{0.6}{\includegraphics{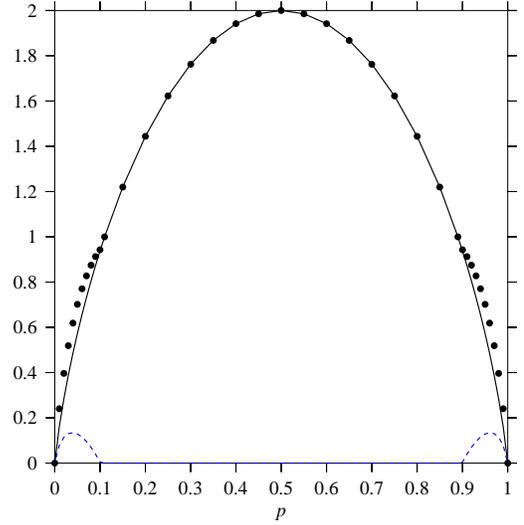}}
\caption{\label{fi:kanc-neq-ksch} Plots of $\Ke(U_p)$ (dots) and
$\Ksch(U_p)$ (solid) as functions of $p$, and of the
difference $\Ke(U_p)-\Ksch(U_p)$ (dashed), demonstrating that
$\Ke(U_p)\ne\Ksch(U_p)$ for some values of $p$.}
\end{figure}
\end{center}

%
%
We show below that \Ke\ is superadditive while \Ksch\ is additive,
which implies that they are not equal for certain unitaries.  We have
also shown this numerically by calculating both functions for a
particular class of unitaries, the Schmidt number 4 family
parametrized by $p$, denoted $U_p$ in Eq.\,(\ref{eq:U_p}).
Fig.\,\ref{fi:kanc-neq-ksch} plots both $\Ke(U_p)$ and $\Ksch(U_p)$ as
a function of $p$, and also their difference.

We now have the tools required to prove that \Ke\ is superadditive, as
promised at the end of the last section.
\begin{theorem} \label{th:superadditivity}
$\Ke$ is superadditive, \ie there exist unitaries $U$ such that
\begin{equation} \label{eq:superadditivity}
\Ke^{\A_1\A_2:\B_1\B_2}(U_{\A_1\B_1}\tp U_{\A_2\B_2})>2\Ke^{\A_1\B_1}
(U_{\A_1\B_1}).
\end{equation}
where the subscripts on $U$ indicate the subsystems to which it is
applied.
\end{theorem}
\proof Let $U=\sqrt{1-p}\,I\tp I+i\sqrt{p}\,X\tp X$.  We show that
additivity is violated for certain values of $p$.  (We will only add
subscripts where necessary.)

Since $U$ has two Schmidt coefficients, Th.\,\ref{th:kanc-eq-ksch}
implies that $\Ke(U)=\Ksch(U)$.  Therefore, the right-hand side of
Eq.\,(\ref{eq:superadditivity}) is $2\Ke(U)=2\Ksch(U)=2H(p)$.

\begin{center}
\begin{figure}[ht]
\scalebox{1.2}{\includegraphics{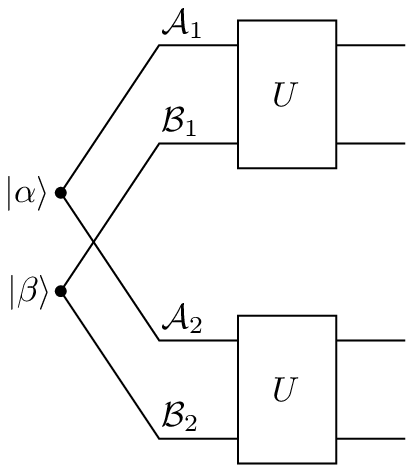}}
\caption{\label{fi:U_tp_U} Diagram of $U\tp U$ applied to systems
$\A_1\B_1$ and $\A_2\B_2$.  Note that $\A_1\A_2$ starts out in the
maximally entangled state $|\alpha\>$, and $\B_1\B_2$ starts out in
the maximally entangled state $|\beta\>$, so $\A_1A_2$ is not
initially entangled with $\B_1\B_2$.}
\end{figure}
\end{center}

To obtain the violation of additivity Eq.\,(\ref{eq:superadditivity})
we now construct specific states $|\alpha\>$ and $|\beta\>$ of \A\ and
\B\ for which $E(U|\alpha\>|\beta\>)>2H(p)$.  To do this, we apply
$U\tp U$ to two pairs of systems, as depicted in
Fig.\,\ref{fi:U_tp_U}, where we have omitted the ancillas as they turn
out not to be necessary for our construction of $|\alpha\>$ and
$|\beta\>$.  Let $|\alpha\>=(|00\>+|11\>)/\sqrt2$ be a state of
Alice's system $\A_1\A_2$ and $|\beta\>=(|00\>+|11\>)/\sqrt2$ be a
state of Bob's system $\B_1\B_2$.

\begin{center}
\begin{figure}[ht]
\scalebox{1.3}{\includegraphics{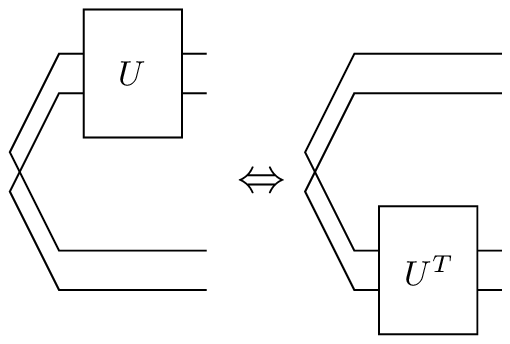}}
\caption{\label{fi:transpose}Illustration of the identity
$U_{\A_1\B_1}\tp I_{\A_2\B_2}|\alpha\>|\beta\>=I_{\A_1\B_1}\tp
U_{\A_2\B_2}^T|\alpha\>|\beta\>$.}
\end{figure}
\end{center}

We make use of a handy identity to calculate $E(U|\alpha\>|\beta\>)$.
Since $|\alpha\>$ and $|\beta\>$ are maximally entangled, a
calculation shows that for any two-qubit unitary $U$,
\begin{equation} \label{eq:transpose-two-qubit}
\(U_{\A_1\B_1}\tp I_{\A_2\B_2}\)|\alpha\>|\beta\>=\(I_{\A_1\B_1}\tp
U_{\A_2\B_2}^T\)|\alpha\>|\beta\>,
\end{equation}
where the transpose is taken in the basis
$\{|00\>,|01\>,|10\>,|11\>\}$.  This is illustrated in
Fig.\,\ref{fi:transpose}.

\begin{center}
\begin{figure}[ht]
\input{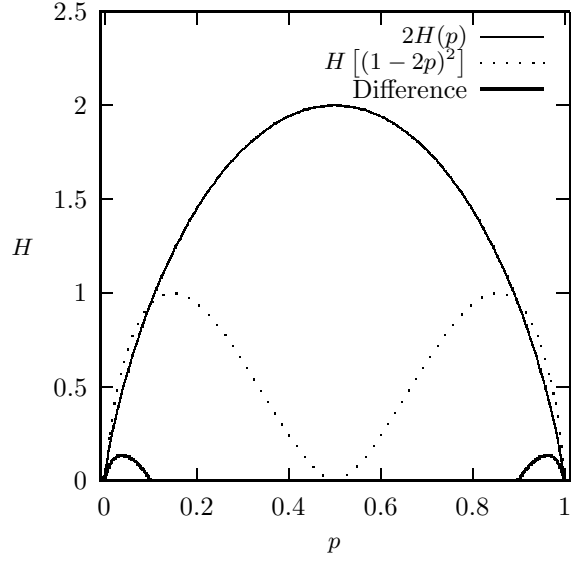}
\caption{\label{fi:h}A plot of $2H(p)$, $H\[(1-2p)^2\]$, and their
difference.}
\end{figure}
\end{center}

For the unitary we are considering, $U^T=U$, so that
Eq.~(\ref{eq:transpose-two-qubit}) implies
\begin{equation}
E(U_{\A_1\B_1}\tp U_{\A_2\B_2}|\alpha\>|\beta\>)=E(I_{\A_1\B_1}\tp
U_{\A_2\B_2}^2|\alpha\>|\beta\>).
\end{equation}
We may now apply Lemma~\ref{le:ksch}, considering $\A_1$ and $\B_1$ as
the ancillas to $\A_2$ and $\B_2$, respectively.  We see that
$E(U_{\A_1\B_1}\tp U_{\A_2\B_2}|\alpha\>|\beta\>)=\Ksch(U^2)$.
Observing that $U^2$ is a unitary with two Schmidt coefficients,
\begin{equation}
U^2=(1-2p)I\tp I+2i\sqrt{p(1-p)}\,X\tp X,
\end{equation}
we obtain
\begin{equation}
\Ke(U\tp U)\ge E(U\tp U|\alpha\>|\beta\>)=H\[(1-2p)^2\],
\end{equation}
so we have reduced the problem to showing that there exist values
of $p$ such that $H\[(1-2p)^2\]>2H(p)$.  The existence of such values
is shown in Fig.\,\ref{fi:h}~\footnote{The similarity
between Fig.\,\ref{fi:kanc-neq-ksch} and Fig.\,\ref{fi:h}
is currently the subject of further
investigation.}.\qed


\subsubsection{Extension to general quantum operations}
\label{subsubse:extension}

Our results to this point have primarily concerned strength measures
for unitary operations.  In this subsection, we obtain some results
for general quantum operations, proving generalizations of
Lemma~\ref{le:ksch} and Th.\,\ref{th:kanc-geq-ksch} to quantum
operations.  We will not do a detailed investigation of the axioms and
properties satisfied by these measures for general operations,
although arguments similar to the unitary case mostly go through.

%
%
The first step is to generalize our definitions of \Ke\ and \Ksch.  In
order to generalize \Ke\ (Eq.\,(\ref{eq:kanc})) to quantum operations,
we must choose an entanglement measure which applies to mixed states
as well as pure states.  We use the {\em entanglement of
  formation}~\cite{Wootters98}:
\begin{equation}
F(\rho)\eq\min\sum\nolimits_j p_j E(|\psi_j\>)
\end{equation}
where the minimization is over all pure state decompositions
$\{p_j,|\psi_j\>\}$ of $\rho$, and $E$ is the entanglement of pure
states.  Note that any two decompositions of $\rho$ are related by a
right unitary matrix $U_{jk}$: $\rho=\sum_j
p_j|\psi_j\>\<\psi_j|=\sum_k q_k|\phi_k\>\<\phi_k|$ if and only
if~\cite{Schrodinger36,Jaynes57b,Hughston93}
$\sqrt{p_j}|\psi_j\>=\sum_j U_{jk}\sqrt{q_k}|\phi_k\>$.  We take as
our generalized $\Ke(\E)$ the maximum entanglement generated by \E\
over all separable input states $\rho_{\A:\B}$:
\begin{equation}
\Ke(\E)=\max_{\rho_{\A:\B}}F\[\E(\rho_{\A:\B})\].
\end{equation}
Note that $F\circ\E$ is a convex function maximized on the convex set
of separable states, $\{\rho_{\A:\B}\}$, and therefore
$F\[\E(\rho_{\A:\B})\]$ achieves its maximum for extreme points of the
set of separable states, \ie pure product states.

To generalize \Ksch, let \E\ be a quantum operation with operation
elements $\{G_k\}$: $\E(\rho)=\sum\nolimits_k G_k\rho G_k^\dag$.
\E\ can be decomposed differently as $\E(\cdot)=\sum_j F_j\cdot
F_j^\dag$ if and only if~\cite{Choi75} the two sets of operation
elements are related by a right unitary matrix: $F_j=\sum_k
U_{jk}G_k$.  By analogy with the entanglement of formation, a natural
definition of $\Ksch(\E)$ is
\begin{equation}
\Ksch(\E)\eq\min\sum_j\frac{\tr(F_j^\dag F_j)}{d_\A d_\B}\Ksch(F_j),
\end{equation}
where $\Ksch(F_j)$ is given by Eq.\,(\ref{eq:op-Ksch}), and the
minimization is over all possible decompositions of \E\ into operation
elements.  The coefficients $\tr(F_j^\dag F_j)/(d_\A d_\B)$ form a
probability distribution.  A physical interpretation is as follows: if
$\Ksch(F_j)$ is the strength of the operation $F_j$, then $\Ksch(\E)$
is the expected strength of \E, minimized over all possible
decompositions of \E.

First, we prove two lemmas generalizing Lemma~\ref{le:ksch}.  For the
remainder of this section, let $|\alpha\>$ be a maximally entangled
state of system \A\ with an ancilla $\R_\A$, and $|\beta\>$ be a
maximally entangled state of \B\ with an ancilla $\R_\B$.

\begin{lemma} \label{lemma:sch-op}
For all operators $Q$,
\begin{equation} \label{eq:op-ksch-ent}
\Ksch(Q)=E\(\textstyle{\sqrt\frac{d_\A d_\B}{\tr(Q^\dag Q)}}Q|\alpha\>
|\beta\>\).
\end{equation}
\end{lemma}
\proof Recall that $\Ksch(Q)=H\(\l\{s_l^2/\tr(Q^\dag Q)\r\}\)$, so we
need only calculate the right-hand side of
Eq.\,(\ref{eq:op-ksch-ent}).  Expand the state $Q|\alpha\>|\beta\>$ as
\begin{equation}
Q|\alpha\>|\beta\>=\sum_l s_l A_l|\alpha\> B_l|\beta\>=\sum_l
\frac{s_l}{\sqrt{d_\A d_\B}}|a_l\>|b_l\>,
\end{equation}
where $\sum_l s_l A_l\tp B_l$ is the Schmidt decomposition for $Q$,
and $|a_l\>\eq\sqrt{d_\A}\,A_l|\alpha\>,|b_l\>\eq\sqrt{d_\B}\,B_l|
\beta\>$ are orthonormal bases for their respective systems.  The
result follows.\qed

\begin{lemma}
  For any quantum operation \E, let
  $\sigma\eq\E(|\alpha\>\<\alpha|\tp|\beta\>\<\beta|)$.  Then
  $\Ksch(\E)=F(\sigma)$, where $F$ is the entanglement of formation.
\end{lemma}
\proof Let $F_j$ be the set of operation elements for \E\ achieving
the minimum in the definition of \Ksch.  Then, applying the definition
and Lemma~\ref{lemma:sch-op}, we have
\begin{equation} \begin{split}
\Ksch(\E)&=\sum_j\frac{\tr(F_j^\dag F_j)}{d_\A d_\B}\Ksch(F_j)\\
&=\sum_j\frac{\tr(F_j^\dag F_j)}{d_\A
d_\B}E\(\textstyle{\sqrt\frac{d_\A d_\B}{\tr(F_j^\dag
F_j)}}F_j|\alpha\>|\beta\>\).
\end{split} \end{equation}
Noting that
\begin{equation}
\l\{\frac{\tr(F_j^\dag F_j)}{d_\A d_\B},\sqrt\frac{d_\A
d_\B}{\tr(F_j^\dag F_j)}F_j|\alpha\>|\beta\>\r\}
\end{equation}
is an ensemble for $\sigma$, we deduce that $\Ksch(\E)\ge F(\sigma)$.
To prove the reverse inequality, suppose $\sigma=\sum_k
p_k|\phi_k\>\<\phi_k|$ is the minimizing decomposition for the
entanglement of formation of $\sigma$.  Note that $\sigma$ can also be
decomposed as
\begin{equation} \label{eq:sigma}
\sigma=\sum\nolimits_j F_j\left(|\alpha\>\<\alpha|\tp|\beta\>\<\beta|
\right)F_j^\dag.
\end{equation}
The minimizing decomposition is related to the decomposition from
Eq.\,(\ref{eq:sigma}) by a right unitary matrix $U$:
$\sqrt{p_k}|\phi_k\>=\sum\nolimits_j U_{kj}F_j|\alpha\>|\beta\>$.
This unitary freedom is identical to the freedom in the operator-sum
decomposition, so the set of elements $G_k=\sum_j U_{kj}F_j$ is also
an operator-sum decomposition for \E, as well as giving the minimizing
decomposition of $\sigma$, that is
$\sqrt{p_k}|\phi_k\>=G_k|\alpha\>|\beta\>$.  This gives us the
desired inequality,
\begin{equation} \begin{split}
F(\sigma)&=\sum_k\frac{\tr(G_k^\dag G_k)}{d_\A
d_\B}E\(\textstyle{\sqrt\frac{d_\A d_\B}{\tr(G_k^\dag
G_k)}}G_k|\alpha\>|\beta\>\)\\
&=\sum_k\frac{\tr(G_k^\dag G_k)}{d_\A d_\B}\Ksch(G_k)\\
&\ge\Ksch(\E).
\end{split} \end{equation}\qed

%
%
The desired bound on \Ke\ now follows:
\begin{theorem}
$\Ke(\E)\ge\Ksch(\E)$ for all quantum operations \E.
\end{theorem}
\proof The result follows immediately from the previous lemma and the
fact that 
\begin{equation}
\Ke(\E)=\max F[\E(\rho_{\A:\B})]\ge F\[\E(|\alpha\>\<\alpha|\tp|\beta
\>\<\beta|)\].
\end{equation}\qed


\subsection{Entanglement generation with prior
entanglement}

In this section we consider the largest change in entanglement which
can be caused by a unitary $U$, using both ancillas and prior
entanglement, as defined in Eq.\,(\ref{eq:kdel}) and repeated here for
convenience: $\Kdel(U)=\sup_{|\psi\>}|E(U|\psi\>)-E(|\psi\>)|$, where
$U$ acts on the combined system \A\B, and $|\psi\>$ is an arbitrary
state of \A\B\ plus their ancillas, $\R_\A$ and $\R_\B$.  We show
that, although \Kdel\ involves a more difficult maximization than \Ke,
and may therefore be more difficult to work with, it satisfies more of
the axioms and properties described in Sec.\,\ref{subse:axiomatic}
than \Ke\ does.  Incidentally, since \Kdel\ and \Ke\ have different
properties they can not, in general, be equal.

%
%
We first show that \Kdel\ obeys the three axioms.  \Kdel\ is clearly
non-negative and satisfies local unitary invariance.  To show that
\Kdel\ satisfies locality is only slightly more involved.  If $U=A\tp
B$, then $\Kdel(A\tp B)=\sup_{|\psi\>}|E(A\tp
B|\psi\>)-E(|\psi\>)|=0$.  On the other hand, since
$\Kdel(U)\ge\Ke(U)$ and we know that $\Ke(U)$ satisfies locality,
$\Kdel(U)=0$ only if $\Ke(U)$, which implies that $U$ is a local
unitary, as required.


Second, we show that \Kdel\ satisfies Properties~\ref{pr:exchange},
\ref{pr:time}, \ref{pr:chaining}, and
\ref{pr:ancilla}--\ref{pr:strong}.  Properties~\ref{pr:exchange}
and~\ref{pr:time}, exchange symmetry and time-reversal invariance, are
easily seen to be true.
%
We do not know whether property~\ref{pr:continuity}, continuity, is
satisfied.  The argument used to establish that $\Ke$ is continuous
does not work in this instance, because we do not have any bound on
the size of the ancilla that \A\ and \B\ may use.  If such a bound
could be established then a similar continuity bound to that used for
$\Ke$ could be proved.  Next, we show that \Kdel\ obeys chaining,
Property~\ref{pr:chaining}.  For any two unitaries $U$ and $V$,
\begin{eqnarray}
&\!\!\!\!\Kdel&\!\!\!\!(UV)=\sup_{|\psi\>}|E(UV|\psi\>)-E(|\psi\>)|
\non\\
&\!\!\!\!=&\!\!\!\!\!\!\sup_{|\psi\>}\l|E(UV|\psi\>)-E(V|\psi\>)+
E(V|\psi\>)-E(|\psi\>)\r|\non\\
&\!\!\!\!\le&\!\!\!\!\!\!\sup_{|\phi\>=V|\psi\>}|E(U|\phi\>)-
E(|\phi\>)|+\sup_{|\psi\>}|E(V|\psi\>)-E(|\psi\>)|\non\\
&\!\!\!\!=&\!\!\!\!\!\!\Kdel(U)+\Kdel(V).
\end{eqnarray}
Property~\ref{pr:ancilla}, stability with respect to ancillas, holds
since \Kdel\ already allows the possibility of arbitrary ancillas.
Therefore, by Proposition~\ref{po:subadditive}, \Kdel\ also satisfies
strong subadditivity, Property~\ref{pr:strong}.  Finally, we note that
the definitions immediately imply that \Kdel\ satisfies the reduction
property, Property~\ref{pr:reduction}.


\section{Strength measures based on metrics} \label{se:metrics}

In this section we consider a class of strength measures motivated by
the axiomatic approach.  This is in contrast to
Sec.\,\ref{se:generation}, where we studied strength measures based on
entanglement generation.  The strength measures we study here are
based on {\em metrics}.  We explore the axioms and properties obeyed
by these measures when different constraints are placed on the
underlying metrics.  We derive an exact, analytic formula for one
particular measure.  Finally, we examine the potential of these
measures for analyzing quantum computational complexity, as described
in Sec.\,\ref{subse:axiomatic}.

Recall the definition of a metric.  Let $\mathbb{X}$ be a set.  A
metric is a real function $D:\mathbb{X}\times\mathbb{X}\to\mathbb{R}$
satisfying the following properties for any three elements $x,y,z$ of
$\mathbb{X}$:
\begin{equation} \begin{split}
&(1)\ D(x,y)\ge0\text{ with equality if and only if }x=y\\
&(2)\ D(x,y)=D(y,x)\ \textbf{(symmetry)}\\
&(3)\ D(x,z)\le D(x,y)+D(y,z)\ \textbf{(triangle inequality)}
\end{split}\non \end{equation}

Given a metric $D$, the corresponding strength measure $\Kd(U)$ is the
minimum distance between $U$ and the set of local unitaries $\LU$:
\begin{equation}
\Kd(U)=\min_{L\in\LU} D(U,L).
\end{equation}
The set \LU\ varies depending on context.  The most common case is
where $U$ is a two-qudit unitary acting on the space \A\B\ and \LU\ is
the set of products of one-qudit unitaries, $\Kd(U)=\min_{A,B}D(U,A\tp
B)$.  Analogues of the definition of \Kd\ were introduced to quantify
state entanglement by Vedral \etal~\cite{Vedral97}, and have been
studied in considerable detail, proving to be a fruitful approach to
quantifying state entanglement.

More generally, if $U$ acts on a composite of systems,
$\A_1,\A_2,\dots,\A_m$, there are several notions of ``local'', which
we differentiate with superscripts.  For example, suppose $U$ acts on
\A\B\C.  One notion of ``local unitary'' corresponds to unitaries of
the form $A\tp B\tp C$, so that
$\Kd^{\A:\B:\C}(U)=\min_{A,B,C}D(U,A\tp B\tp C)$.  A different
division into subsystems leads to a different measure:
$\Kd^{\A:\B\C}(U)=\min_{A,B}D(U,A\tp B)$, where $A$ acts on system \A\ 
but now $B$ is any unitary on \B\C.



\subsection{Properties of strength measures based on metrics }

One reason for studying strength measures based on metrics is that the
properties of the strength measure may be controlled by varying the
properties of the underlying metric.  We consider strength measures
based on: (1) arbitrary metrics; (2) metrics invariant under local
unitaries; and (3) metrics invariant under any unitary.  Each extra
requirement causes the strength measure to obey extra axioms and
properties from Sec.\,\ref{subse:axiomatic}.  Since we know of no
general way to characterize families of metrics, in this section we do
not consider any of the properties applying to families
(Properties~\ref{pr:systems}--\ref{pr:reduction}).  Therefore,
throughout this section we assume $\Kd=\Kd^{\A:\B}$.

The metric properties are easily seen to guarantee that the axioms of
non-negativity and locality hold for all \Kd.  An elegant fact is that
the metric properties alone also imply that \Kd\ satisfies the continuity
property:
\begin{lemma}
For any two unitaries $U$ and $V$, and any metric $D$,
$|\Kd(U)-\Kd(V)|\le D(U,V)$.
\end{lemma}
\proof Choose $A$ and $B$ such that $\Kd(V)=D(V,A\tp B)$.  By
definition $\Kd(U)\le D(U,A\tp B)$, and by the triangle inequality
$D(U,A\tp B)\le D(U,V)+D(V,A\tp B)=D(U,V)+\Kd(V)$.  Thus $\Kd(U)\le
D(U,V)+\Kd(V)$, which may be rearranged to give $\Kd(U)-\Kd(V)\le
D(U,V)$.  By symmetry, $\Kd(V)-\Kd(U)\le D(U,V)$.\qed

If $D$ is locally unitarily invariant, \ie, $D(U,V)=D[(A\tp B)U,(A\tp
B)V]=D[U(A\tp B),V(A\tp B)]$, then \Kd satisfies local unitary
invariance.

Finally, suppose the metric satisfies full unitary invariance, so that
$D(U,V)=D(WU,WV)=D(UW,VW)$ for any unitary $W$.  Then \Kd\ satisfies
two additional properties.  The first is exchange symmetry, which is
easily proved.
The second is chaining, $\Kd(UV)\le\Kd(U)+\Kd(V)$.  To see this,
suppose $A\tp B$ and $C\tp D$ minimize $\Kd(U)$ and $\Kd(V)$,
respectively.  Then
\begin{eqnarray}
\!\!&\Kd&\!\!\!\!(UV)\le D\[UV,(A\tp B)(C\tp D)\]\non\\
&\le&\!\!\!\!D\[UV,U(C\tp D)\]+D\[U(C\tp D),(A\tp B)(C\tp D)\]\non\\
&=&\!\!\!\!D(V,C\tp D)+D(U,A\tp B)\non\\
&=&\!\!\!\!\Kd(U)+\Kd(V).
\end{eqnarray}



\subsection{An explicit formula for the Hilbert-Schmidt strength of a
two-qubit unitary}

%
%
In this section we consider an example of a metric-based strength
measure, the {\em Hilbert-Schmidt strength} \Khs\ induced by the
unitarily invariant {\em Hilbert-Schmidt} norm on operators,
$\|Q\|_{{\rm HS}}\eq\sqrt{\tr(Q^\dag Q)}$.  More explicitly, for a
bipartite unitary operation $U$ we define
\begin{equation}
\Khs(U)\eq\min_{A,B}\|U-A\tp B\|_{{\rm HS}},
\end{equation}
where $A$ and $B$ are local unitary operators on the respective
subsystems.  We now exhibit an explicit formula for the
Hilbert-Schmidt strength in the two-qubit case.

%
%
The statement of the result is simplified by first making some
definitions and observations.  Let $U$ be a two-qubit unitary
operation with canonical decomposition
\begin{equation} \label{eq:HS-can-decomp}
U= (A_1\tp B_1)e^{i(\theta_1 X\tp X+\theta_2 Y\tp Y+\theta_3 Z\tp
Z)}(A_2\tp B_2).
\end{equation}
Because of local unitary invariance the Hilbert-Schmidt strength
depends only on the parameters $\theta_j$, that is, we can ignore the
local unitary operations $A_{1,2}$ and $B_{1,2}$.  Without loss of
generality, we assume $U$ is in {\em canonical form}, that is,
$A_1=B_1=A_2=B_2=I$.

%
%
We define $|\phi_0\>=(|00\>+|11\>)/\sqrt2$, and
$|\phi_j\>\eq(I\tp\sigma_j)|\phi_0\>$ for $j=1,2,3$ where we write
$\sigma_0,\sigma_1,\sigma_2,\sigma_3$ to denote $I,X,Y,Z$.  Note that
the set $|\phi_j\>$ for $j=0,1,2,3$ is the Bell basis, up to phases.
A simple but tedious calculation verifies the useful formula
$\<\phi_j|\sigma_k\tp\sigma_l|\phi_j\>=\delta_{kl}H_{jk}$, where the
$4\times4$ matrix $H$ is
\begin{equation}
H=\[\begin{smallmatrix}
1&\ \ 1&-1&\ \ 1\non\\
1&\ \ 1&\ \;1&-1\non\\
1&-1&-1&-1\non\\
1&-1&\ \;1&\ \ 1
\end{smallmatrix}\].
\end{equation}

%
%
The $H$ matrix can also be used to evaluate the eigenvalues of $U$.
Because $X\tp X$, $Y\tp Y$, and $Z\tp Z$ are diagonal in the
$|\phi_j\>$ basis, $U$ may be written in diagonal form as
$U=\sum_j\lambda_j|\phi_j\>\<\phi_j|$, where $\lambda_j$ are the
eigenvalues of $U$.  These eigenvalues are evaluated as follows:
\begin{equation} \begin{split}
\lambda_j&=\<\phi_j|U|\phi_j\>\\
&=\left\<\phi_j\left|e^{i(\theta_1 X\tp
X+\theta_2 Y\tp Y+\theta_3 Z\tp Z)}\right|\phi_j\right\>\\
&=\exp\(i\sum_{k=1}^3\theta_k\<\phi_j|\sigma_k\tp\sigma_k|\phi_j\>\),
\end{split} \end{equation}
where in the last line we used the fact that all three
$\sigma_k\tp\sigma_k$ are diagonal in the $|\phi_j\>$ basis.
Substituting $\<\phi_j|\sigma_k\tp\sigma_l|\phi_j\>=\delta_{kl}H_{jk}$
we obtain:
\begin{equation}
\lambda_j=\exp\(i\sum_{k=1}^3 H_{jk}\theta_k\).
\end{equation}

%
%
\begin{theorem} For a two-qubit unitary $U$ with canonical
decomposition Eq.\,(\ref{eq:HS-can-decomp}), the Hilbert-Schmidt
strength is given by the formula
\begin{equation}
\Khs(U)=\sqrt{8-2\max_{0\le k\le3}\textstyle{\l|\sum_j\lambda_j
H_{jk}\r|}}.
\end{equation}
The minimizing local unitary is $A\tp
B=e^{i\theta}\sigma_k\tp\sigma_k$ where $k$ achieves the maximum in
the expression above, and $\theta$ is the argument of
$\sum\nolimits_j\lambda_jH_{jk}$.
\end{theorem}

\proof Simple algebra shows that
\begin{equation}
\Khs(U)^2=\min_{A,B}\[8-2\Re\(\tr\[U^\dag(A\tp B)\]\)\],
\end{equation}
where $\Re(\cdot)$ denotes the real part.  We expand $A$ and $B$ in
terms of the Pauli operators as $A=\sum_{k=0}^3 a_k\sigma_k$,
$B=\sum_{l=0}^3 b_l\sigma_l$.  (Note that the unitarity of $A$ and $B$
implies that $\sum_k|a_k|^2=\sum_l|b_l|^2=1$.)  Substituting these
expressions for $A$ and $B$, and
$U=\sum_j\lambda_j|\phi_j\>\<\phi_j|$, gives
\begin{equation}
\Khs(U)^2=\min_{a_k,b_l}\[8-2\Re\(\sum_{jkl}\lambda_j^* a_k
b_l\<\phi_j| \sigma_k\tp\sigma_l|\phi_j\>\)\],
\end{equation}
where the minimization is over all $a_k,b_l$ such that the
corresponding $A$ and $B$ are unitary.  But
$\<\phi_j|\sigma_k\tp\sigma_l|\phi_j\>\>=\delta_{kl}H_{jk}$, as noted
earlier, so this expression simplifies to
\begin{equation}
\Khs(U)^2=8-2\max_{a_k,b_k}\Re\(\textstyle{\sum_{jk}\lambda_j^* a_k b_k H_{jk}
} \).
\end{equation}
The Cauchy-Schwarz inequality implies $\sum_k|a_k b_k|\le1$, so:
\begin{eqnarray}
\Re\(\textstyle{
\sum_{jk}\lambda_j^* a_k b_k H_{jk} }
\)\!\!\!\!\!\!&&\le\sum
\nolimits_k\(|a_k b_k|\l|\sum\nolimits_j\lambda_j^* H_{jk}\r|\)\non\\
&&\le\max_k\l|\sum\nolimits_j\lambda_j^* H_{jk}\r|\non\\
&&=\max_k\l|\sum\nolimits_j\lambda_j H_{jk}\r|.
\end{eqnarray}
Equality occurs when $a_l=\delta_{kl}$ and
$b_l=\delta_{kl}e^{i\theta}$, where $k$ maximizes the right-hand side
of the inequality, and $e^{i\theta}\sum\nolimits_j\lambda_j^*
H_{jk}=\l|\sum\nolimits_j\lambda_j H_{jk}\r|$.  This corresponds to
$A\tp B=e^{i\theta}\sigma_k\tp\sigma_k$, and $\theta$ as described in
the statement of the theorem.\qed


\subsection{Applications to computational complexity}

We have seen that strength measures based on unitarily invariant
metrics satisfy many desirable axioms and properties.  It is natural
to ask whether these measures might be useful in answering questions
about computational complexity, as described in
Sec.~\ref{subse:axiomatic}.

In order for a family of measures \{\Kd\} to be useful in this
context, we require \{\Kd\} to be stable under addition of systems
(for the remainder of this section, we simply write ``stable'' for
this property).  This is to ensure that the strength of a \cnot\ gate
is independent of the number of qubits in the system being studied.
It is tempting to consider a family of measures \{\Kd\} whose
underlying family of metrics is stable, in the sense that
$D(U,V)=D(U\tp I,V\tp I)$ for any unitaries $U$ and $V$.  However, we
show here that such metrics give rise to trivial bounds on
computational complexity.  Denote by $U$ a unitary acting on $n$
qubits, and let $0$ and $I$ be the zero and identity operator,
respectively, on $n$ qubits.  For any such unitary,
\begin{eqnarray}
\Kd(U)&= & \min_{A_1,\ldots,A_n}D(U,A_1\tp\ldots\tp A_n) \nonumber \\
&\le & \min_{A_1,\ldots,A_n}\[D(U,0)+D(0,A_1\tp\ldots\tp A_n)\] \nonumber \\
&= & 2D(I,0),
\end{eqnarray}
where to obtain the last line we used the unitary invariance of $D$.
But $I=I_1\tp I_2\tp\cdots\tp I_n$, where $I_j$ is the identity on the
$j$th qubit, so by the metric stability property $\Kd(U)$ is always
bounded by $2D(I,0)=2D(I_1,0)$, which is a constant.  Therefore, the
lower bound on the number of two-qubit gates required to implement a
$n$-qubit gate, $k\ge K(U)/\Kmax$ (Eq.\,(\ref{eq:lower-bound})), is a
constant.

This shows that any family of metrics which is both unitarily
invariant and stable cannot give interesting lower bounds on
computational complexity.  As noted above, unitary invariance is a
useful property.  On the other hand, stability of the family of
metrics may not be necessary for stability of the induced family of
measures.  So, it may be possible to find a family of unitarily
invariant metrics which is not stable, but which induces a stable
family of measures, and could therefore give useful lower bounds on
computational complexity.

\begin{table}[t] 
\begin{tabular}{|l|ccccccc|}
\hline
Measure:                   &\Khar&\Ksch&\Ke&\Kdel&\Kd&\Kd[LU]&\Kd[U]\\
\hline
A1 \ssize Non-negativity   &yes  &yes  &yes&yes  &yes&yes    &yes   \\
A2 \ssize Locality         &yes  &yes  &yes&yes  &yes&yes    &yes   \\
A3 \ssize LU invariance    &yes  &yes  &yes&yes  &?  &yes    &yes   \\
P1 \ssize Exchange         &yes  &yes  &yes&yes  &?  &?      &yes   \\
P2 \ssize Time-reversal    &yes  &yes  &?  &yes  &?  &?      &?     \\
P3 \ssize Continuity       &no   &yes  &yes&?    &yes&yes    &yes   \\
P4 \ssize Chaining         &yes  &no   &no &yes  &?  &?      &yes   \\
P5 \ssize System stability &--   &--   &-- &--   &?  &?      &?     \\
P6 \ssize Ancilla stability&yes  &yes  &yes&yes  &?  &?      &?     \\
P7 \ssize Weak additivity  &yes  &yes  &no &yes  &?  &?      &?     \\
P8 \ssize Strong additivity&yes  &yes  &no &yes  &?  &?      &?     \\
P9 \ssize Reduction        &yes  &no   &yes&yes  &?  &?      &?     \\
\hline
\end{tabular}
\caption{\label{ta:summary}Summary of axioms and properties of
strength measures.  ``yes''/''no'' indicates whether the
strength measure obeys the axiom/property.  ``--''
means the property is
not applicable, and ``?''  means we do not know whether the
strength measure obeys the axiom/property.  \Kd[LU] refers to strength
measures induced by locally unitarily invariant metrics, and \Kd[U]
refers to strength measures induced by unitarily invariant metrics.}
\end{table}


\section{Summary and future directions} \label{se:future}

We have developed the beginnings of a quantitative theory of quantum
dynamical operations as a physical resource. While promising
preliminary results have been obtained, an enormous amount of work
remains to be done. (Table~\ref{ta:summary} summarizes the properties
of the strength measures we investigated.)  We believe the development
of this theory offers a concrete path to address the fundamental
question of quantum computational complexity: how many one- and
two-qubit quantum operations are required to do some desired quantum
operation \E?  This will, in turn, allow us to answer questions about
the relationship of quantum and classical complexity classes, and may
enable the resolution of some
longstanding questions in complexity theory.\\

\noindent{\bf Acknowledgments} 
We thank Charlene Ahn, Sean Barrett, Tony Bracken, Andrew Childs, and
Xiaoguang Wang for helpful discussions.  MAN thanks Raymond Laflamme
for an enjoyable 1998 discussion about the idea of quantifying
the ``entangling power'' of a quantum dynamical operation.  AWH
acknowledges support from the NSA and ARDA under ARO contract no.\ 
DAAD19-01-1-06, and thanks the other authors for their hospitality.

\bibliographystyle{apsrev}

\end{document}